\documentclass{article}

\newcommand{\bt}{\begin{tabular}{c}}
\newcommand{\et}{\end{tabular}}

\newcommand{\eb}{\ee\be } 
\newcommand{\ebp}{\rt.\ee\be\lt.} 
\newcommand{\bmat}{\lt ( \begin{array} }
\newcommand{\emat}{  \end{array} \rt )}

\newcommand{\ED}{\end{document}}

\newcommand{\oy}{{\ov \y}}

\newcommand{\oC}{{\ov C}}

\newcommand{\oF}{{\ov F}}
\newcommand{\A}{{\ov A}}

\renewcommand{\a}{\alpha}	
\renewcommand{\b}{\beta}
\newcommand{\g}{\gamma}
\renewcommand{\d}{\delta}
\newcommand{\e}{\epsilon}
\newcommand{\ve}{\varepsilon}
\newcommand{\z}{\zeta}
\newcommand{\h}{\eta}
\newcommand{\q}{\theta}

\renewcommand{\l}{\lambda}
\newcommand{\m}{\mu}
\newcommand{\n}{\nu}	
\newcommand{\x}{\xi}
	
\renewcommand{\r}{\rho}
\newcommand{\s}{\sigma}

\newcommand{\f}{\phi}
\renewcommand{\c}{\chi}
\newcommand{\y}{\psi}
\newcommand{\w}{\omega}
\newcommand{\G}{\Gamma}
\newcommand{\D}{\Delta}
	
\renewcommand{\L}{\Lambda}
\newcommand{\X}{\Xi}
\renewcommand{\P}{\Pi}

\newcommand{\la}{\label}
\newcommand{\ci}{\cite}

\newcommand{\ds}{\documentstyle}	
\newcommand{\fr}{\frac}

\newcommand{\pa}{\partial}
\newcommand{\ov}{\overline}
\newcommand{\be}{\begin{equation}}
\newcommand{\ee}{\end{equation}}
\newcommand{\ba}{\begin{array}} 
\newcommand{\ea}{\end{array}}
\newcommand{\bea}{\begin{eqnarray}}
\newcommand{\eea}{\end{eqnarray}}
\newcommand{\ra}{\rightarrow}
\newcommand{\Ra}{\Rightarrow}

\newcommand{\lt}{\left}
\newcommand{\rt}{\right}

\newcommand{\ben}{\begin{enumerate}}
\newcommand{\een}{\end{enumerate}}
\newcommand{\bitem}{\begin{itemize}}
\newcommand{\eitem}{\end{itemize}}
\newcommand{\articlenumber}{}

\newcommand{\articletitle}{ 
 Introduction to the BRS Cohomology \\of the Massless Wess Zumino Model:\\ Cybersusy II }

\begin{document}
\makeatletter	   
\renewcommand{\ps@plain}{%
\renewcommand{\@oddhead}{{\articlenumber  \hspace{1cm} }\hspace{1cm}  \hfil\textrm{\thepage}} 
\renewcommand{\@evenhead}{\@oddhead}
\renewcommand{\@oddfoot}{\textrm{\articlenumber \hspace{1cm}  }\hspace{1cm} \hfil\textrm{\thepage}}
\renewcommand{\@evenfoot}{\@oddfoot}}
\makeatother    
\title{  \articletitle \\ \articlenumber}
\author{ J. A.  Dixon\footnote{jadix@telus.net}\\ Dixon Law Firm\footnote{Fax: (403) 266-1487} \\1020 Canadian Centre\\
833 - 4th Ave. S. W. \\ Calgary, Alberta \\ Canada T2P 3T5 }
\maketitle
\pagestyle{plain}

\abstract{\large This paper is the second paper in a series of four papers that introduce cybersusy, which is a new method for analyzing supersymmetry breaking in the standard supersymmetric model (SSM).  The first paper was a summary of the results and the three next papers set out the details. 
In this second paper, we derive the full BRS operator  and action for the general massless Wess-Zumino chiral supersymmetry action.  This includes the source terms which bring in the equations of motion.  The auxiliary field is integrated, which removes manifest supersymmetry, but which allows the Legendre transform to operate correctly to define one-particle-irreducible vertices from the connected Green's functions.  Then  some special terms in the BRS cohomology   are described, together with the constraint equations that they must satisfy. These  `simple dotspinors'  are generated by a `fundamental  dotspinor', which is constructed partly from the Zinn sources. The equations of motion play a very important role in the cohomology for this theory.  These dotspinors play an interesting role in the BRS cohomology of the standard model, which is the subject of the third paper in the series.}
\Large

\section{Introduction}

In the previous paper \ci{cybersusyI}, which was the first of a series of four papers, a summary of a new mechanism for supersymmetry breaking in the SSM was outlined, and the results were summarized for the leptons. 

The mechanism was based on the cybersusy algebra which arises for composite operators in the BRS cohomology of the SSM when gauge symmetry is broken.  When this algebra is used to construct an effective action with the appropriate effective fields, supersymmetry is explicitly broken in a unique way.  This breaking occurs sector by sector for each set of different quantum numbers.  In \ci{cybersusyI} we looked at the leptons.  The various kinds of baryons also look promising, as we shall see in 
\ci{cybersusyIII}, although the masses are not yet worked out for any of the examples for that case.

In  \ci{cybersusyI}, we looked at only some small parts of a few composite operators, but the mechanism is much more general than that, as will be seen in \ci{cybersusyIII}.  But first we need more machinery to look at the composite operators.

This is the second paper of the series. This paper presents the derivation of the BRS nilpotent operator $\d$ for the massless but interacting chiral scalar Wess-Zumino action, but it does so in rather a special way.  The main feature of this derivation is that the auxiliary fields $F^i$ are integrated out, so that manifest supersymmetry is lost.  There are several reasons for doing this:
\ben
\item
The auxiliary field $F^i$ has no momentum dependent kinetic term, and so is non-propagating.  This means that it does not fit into the usual scheme for reducing the connected diagrams into one-particle-irreducible diagrams using the Legendre transform.
This is an essential part of the formulation of the BRS identity using the sources of Zinn-Justin. 
\item
If one keeps the auxiliary field without integrating it, then the cohomology will be dependent on  $F^i$.  There is then an issue about integrating the auxiliary field in the cohomology, which is again hard to sort out.
\een 

In summary, no error can be introduced by integrating the auxiliary field
$F^i$ , whereas keeping it unintegrated creates numerous puzzles that seem hard to sort out. Manifest supersymmetry is lost by doing this, but supersymmetry is still present, albeit somewhat more obscurely.  However the supersymmetry is embedded in the BRS operator and its nilpotence, as we shall show. 

The integration results in constraints and a whole new set of invariants that are not apparent when superfields are used. 

The massless interacting theory is chosen because we are interested in applying these results to the massless standard supersymmetric model in the third paper \ci{cybersusyIII}.

We will explain how to pick out some special terms in the BRS cohomology of
 $\d_{\rm BRS}$.
This special set is an infinite  set of ghost charge zero composite operators that transform under $\d_{\rm BRS}$  as though they were chiral dotted spinor superfields. 
We will call this infinite  set the `simple dotspinors'. 

 They are all generated by a `fundamental dotspinor'. However the fundamental dotspinor does {\bf not} transform as a superfield.  It has extra inhomogeneous terms in its transformation. In composite operators including the fundamental dotspinor, it is possible to introduce constraints so that the composite transforms like a superfield, even though the constituents do not.  To be specific, we will find that there is a scalar {\bf partly composite} operator that transforms exactly like a superfield:
\be
\d {\hat A}^{i} =  ( C\cdot Q +\ov C\cdot \ov Q ) {\hat A}^{i} 
\equiv \d_{\rm SS} {\hat A}^{i}  \ee
where $\d_{\rm SS} $ is what one expects for the transformation of a chiral scalar superfield.  The fundamental dotspinor, however, transforms like this:
\be
\d {\hat \f}_{i \dot \a} =  ( C\cdot Q +\ov C\cdot \ov Q ) {\hat \f}_{i \dot \a}   -
 g_{ijk} 
{\hat A}^j {\hat A}^k \ov C_{\dot \a}
\eb
\equiv \d_{\rm SS}  {\hat \f}_{i \dot \a}
 -
 g_{ijk} 
{\hat A}^j {\hat A}^k \ov C_{\dot \a}
\la{gqwggrgqerge}
\ee
The quadratic inhomogeneous part $ g_{ijk} 
{\hat A}^j {\hat A}^k \ov C_{\dot \a}$ in 
(\ref{gqwggrgqerge}) drops out of the transformation of certain symmetric composites formed from 
products of ${\hat A}^{i}$ and ${\hat \f}_{j \dot \a}$, provided that:
\be
 {g}_{s (j_{n+1} j_{n+2}}   
f_{ j_1 \cdots j_{n})}^{[s i_2 \cdots i_{2m+1}]}
=0
\la{qfggrtegrtegege}
\ee
where the expression must be symmetrized over the indices $(j_1 \cdots j_{n} j_{n+1} j_{n+2} )$, and where the tensor $f_{( j_1 \cdots j_{n})}^{[i_1 i_2 \cdots i_{2m+1}]}$ is used to put together the product of factors of ${\hat A}^{j_r}$ and ${\hat \f}_{i_s \dot \a_s}$.  This will be shown in detail below. These rather peculiar results were found using spectral sequences 
\ci{bigpaper}. However they will be derived here in a more explicit and usable way, without reference to their origin. 
There are two reasons for this
\ben
\item
The spectral sequence is long, hard and has unsolved problems.
\item
We need the explicit results anyway, and the spectral sequence does not give them.
\een
The derivation of equation (\ref{gqwggrgqerge}) is done very explicitly below.  It is a curious fact that there are two ways to get equation (\ref{gqwggrgqerge}) (spectral sequence, and detailed component calculation), and both of them are quite arduous, although the result is a simple one.  Equation (\ref{qfggrtegrtegege})
 follows easily from equation (\ref{gqwggrgqerge}), once one has made appropriate definitions.

In the next paper   of this series \ci{cybersusyIII}, these results will be applied to the supersymmetric standard model (SSM).

\section{The Wess Zumino model and its cohomology}

First we will derive the form of $\d_{\rm BRS}$ for this model.

\section{BRS Transformations with Zinn Sources for the Wess Zumino Model}

\subsection{Chiral Supersymmetry Transformations}

In Table \ref{Purechiral1}, we summarize the Field Transformations  for Pure Chiral Supersymmetry \ci{WZ}.  These transformations close in the sense that
\be
\d^2 = 0.
\ee

The quantity $C_{\a}$, its complex conjugate $\ov C_{\dot \a}$,   and $\x_{\a \dot \b}$ are space-time constant supersymmetry ghosts. $C_{\a}$ is a commuting quantity and  $\x_{\a \dot \b}$ is anticommuting. When this theory is embedded in supergravity, these become space-time dependent, but we shall not consider that here.  There is plenty going on in the rigid supersymmetric theory to occupy our attention for the time being.

\be
\la{Purechiral1}
\begin{array}{|ccc|}  
\hline
\multicolumn{3}{|c|}{\rm   Chiral \; Transformations }
\\
\hline
\d A^i&= & 
  \y^{i}_{  \b} {C}^{  \b} 
+ \x^{\g \dot \d} \partial_{\g \dot \d} A^i
\\
\d {\ov A}_i&= & 
 {\ov \y}_{i  \dot \b} {\ov C}^{ \dot  \b} 
+ \x^{\g \dot \d} \partial_{\g \dot \d} {\ov A}_i
\\
 \d \y_{\a}^i &  =& 
\pa_{ \a \dot \b }  A^{i} {\ov C}^{\dot \b}  
+ 
C_{\a}   
F^i
+ \x^{\g \dot \d} \partial_{\g \dot \d}  \y^{i}_{\a  }
\\
\d
 {\ov \y}_{i \dot \a} &  =& 
\pa_{ \a \dot \a }  {\ov A}_{i} {C}^{\a}  
+ 
{\ov C}_{\dot \a}   
{\ov F}_{i}
+ \x^{\g \dot \d} \partial_{\g \dot \d} 
 {\ov \y}_{i \dot \a} 
\\ 
\d F^i &  =&  \pa_{ \a \dot \b }  \y^{i \a}  {\ov C}^{\dot \b}  
 + \x^{\g \dot \d} \partial_{\g \dot \d}  F^i
\\ 
\d \oF_i &  =&  \pa_{ \a \dot \b }  \oy^{\dot \b}_i  C^{\a}
 + \x^{\g \dot \d} \partial_{\g \dot \d}  \oF_i
\\
\d \x_{\a \dot \b} &  =& - C_{\a} \oC_{\dot \b}
\\
\hline
\end{array}
\ee

\subsection{Action and BRS Identity for massive interacting chiral supersymmetry including Zinn-Justin's sources}

\la{brssection}

The most obvious way to formulate the BRS-ZJ identity is to start with the following action 
\ci{WZ} \ci{ZJ}.

\be
{\cal A}_{\rm Total} 
=
{\cal A}_{\rm WZ} 
+
{\cal A}_{\rm ZJ} 
+
{\cal A}_{\rm Sources} 
\ee
where we will start with the massless Wess-Zumino action, which is \ci{WZ}:
\[
{\cal A}_{\rm WZ} 
= \int d^4 x \;\lt \{
F^i \oF_i   -
\y^{i }_{\a  }     \pa^{\a \dot \b  }   
{\ov \y}_{i \dot \b}
+
\fr{1}{2} 
\pa_{ \a \dot \b  }   A^{i}    \pa^{\a \dot \b  } {\ov  A}_{k} 
\rt.
\]
\be
\lt.
+
g_{ijk}\lt (
   F^{i}   A^j A^k    -    \y^{i \a} \y^j_{\a} A^k   
\rt )  
+
{\ov g}^{ijk}\lt (
   \ov F_{i}   \ov A_j \ov A_k    -    \oy_{i}^{\dot  \a} \oy_{j \dot  \a}  \ov A_k   
\rt ) 
\rt \}
\ee
Using the BRS operator $\d$ defined in equation   
(\ref{Purechiral1}), this action satisfies the invariance 
\be
\d {\cal A}_{\rm WZ} =0
\ee
and the operator $\d$ satisfies the nilpotence condition
\be \d^2 =0
\ee

  The Zinn Justin action  \ci{ZJ} is formed from sources coupled to the variations in equation (\ref{Purechiral1}):
\[
{\cal A}_{\rm ZJ} 
= \int d^4  x\;
\]
\[
\lt \{
 \G_i    \y^{i}_{  \b} { C}^{  \b} 
+
 Y_{i}^{ \a}
\lt (
   \pa_{ \a \dot \b }  A^{i} {\ov C}^{\dot \b}   
+
   F^{i} C_{\a}
\rt )
\rt.
\]
\[
\lt.+
\x^{\g \dot \d} \lt ( 
 Y_{i}^{ \a}  \pa_{\g \dot \d}  \y^{i}_{ \a}
-
 \G_{i}   \pa_{\g \dot \d}  A^{i} 
\rt ) 
+   {\rm Complex\; Conjugate} \rt \}
\]
\be
-
X^{\g \dot \d} C_{\g} {\ov C}_{\dot \d} 
\ee
and the Source term is:
\be
{\cal A}_{\rm Sources} 
=
\int d^4 x
\left \{
{\breve A}_i A^i
+ {\breve \y^{\a}}_i
\y_{\a}^i
+   {\rm Complex\; Conjugate}
 \right \}
\ee
We do not introduce sources for the auxiliary field $F$ or for its variation, because we will  integrate the auxiliary field while we formulate the BRS identity.

\subsection{`Physical' Formulation of BRS-ZJ Identity}

\la{physzj} 

As usual one defines a set of Green's functions by an integral over paths: 
\be
{\cal G}_{\rm Disconnected.} =
e^{i {\cal G}_{\rm Connected.} } =
\ee
\be
\la{disconn2}
\P_{x} 
\int \;
\d A^i
\d {\ov A}_i
\d \y_{\a}^i
\d {\ov \y}_{i \dot \a}
\d F^i
\d {\ov F}_i
e^{i 
{\cal A}_{ \rm  Total } }
\ee

Now we make  the field transformations (and their complex conjugates):
\be
A^i \ra A^i + \ve \d A^i
\ee
\be
\y_{\a}^i \ra \y_{\a}^i + \ve \d \y_{\a}^i
\ee
\be
F^i \ra F^i + \ve \d F^i
\ee
where $\e$ is an anticommuting quantity and $\d$ is defined by equation (\ref{Purechiral1}).  Using the invariance of the action under this field transformation yields the identity:
\[
\int d^4 x
\left \{
{\breve A}_i \fr{\d {\cal G}_{\rm Connected}}{\d \G_i} 
- {\breve \y^{\a}}_i
\fr{\d {\cal G}_{\rm Connected}}{\d  Y^{\a}_i}
\rt.
\]
\[
\lt.
+
{\breve {\ov A}}^i
 \fr{\d {\cal G}_{\rm Connected}}{\d {\ov \G}^i }
 - {\breve {\ov \y}}^{i \dot \a}
 \fr{\d {\cal G}_{\rm Connected}}{\d  {\ov Y}^{i \dot \a}}
\right \}
\]
\be
\la{correctone} + C^{\a} {\ov C}^{ \dot \b}   \fr{\pa {\cal G}_{\rm Connected}}{\pa  \x^{\a \dot \b} }
 = 0
\ee

\subsection{Action for massive interacting chiral supersymmetry after integration of auxiliary $F^i$}

By performing the integration of $F^i$ and $\oF_i$, which can be done by completing the square since there is no kinetic term for the auxiliary fields, this can be written:

\be
{\cal G}_{\rm Disconnected.} =
e^{i {\cal G}_{\rm Connected.} } =
\ee
\be
\la{disconn3}
=
\P_{x} 
\int \;
\d A^i
\d {\ov A}_i
\d \y_{\a}^i
\d {\ov \y}_{i \dot \a}
e^{i \lt \{
{\cal A}_{\rm Physical }+ {\cal A}_{\rm Sources}
\rt \} }
\ee
where
\[
{\cal A}_{\rm Physical}
\]
\[
= \int d^4 x \;\lt \{
- \lt (    
 {\ov g}^{ijk} {\ov A}_j  {\ov A}_k  +
{\ov Y}^{i \dot \b} {\ov C}_{\dot \b } 
\rt )
\rt.
\]
\[
\lt (
{g}_{ilq} A^l A^q
+
{Y}_i^{  \b} {c}_{\b } 
\rt )
  -
\y^{i }_{\a  }     \pa^{\a \dot \b  }   
{\ov \y}_{i \dot \b}
\]
\[
  - g_{ijk}  \y^{i \a} \y^j_{\a} A^k   
  -
 {\ov g}^{ijk} {\ov \y}_{i }^{\dot \a} {\ov \y}_{j  \dot \a} {\ov A}_k 
\]
\[
+
\fr{1}{2} 
\pa_{ \a \dot \b  }   A^{i}    \pa^{\a \dot \b  } {\ov  A}_{k} 
+
\G_i    \y^{i}_{  \b} {  c}^{  \b} 
+
{\ov \G}^i  {\ov \y}_{i \dot  \b} {\ov C}^{\dot   \b}  
\]
\[
+
 Y_{i}^{ \a}   \pa_{ \a \dot \b }  A^{j} {\ov C}^{\dot \b}   
+
 \ov{Y}^{i \dot \b}  \pa_{ \a \dot \b }  {\ov A}_{j} 
{  c}^{  \a}  
\]
\[
\lt.
+
\x^{\g \dot \d} \lt ( 
 Y_{i}^{ \a}  \pa_{\g \dot \d}  \y^{i}_{ \a}
+
 {\ov Y}^{i \dot \b}  \pa_{\g \dot \d}{\ov \y}_{i \dot \b}  
-
 \G_{i}   \pa_{\g \dot \d}  A^{i} 
-
 {\ov \G}^{i  }  \pa_{\g \dot \d}{\ov A}_{i }  
\rt )
\rt \}
\]
\be
-
X^{\g \dot \d} C_{\g} {\ov C}_{\dot \d} 
\la{Aphysical}
\ee

\subsection{BRS-ZJ identity in the Physical Formulation }

\la{brsinphyspform}

A Legendre transform now takes the connected Green's functional to the 1PI functional.   The Legendre transform is of the form:

\be
{\cal G}_{\rm Connected.} =
{\cal G}_{\rm 1PI} +
\int d^4 x
\left \{
{\breve A}_i A^i
+
 {\breve \y^{\a}}_i
\y_{\a}^i
+   {\rm Complex\; Conjugate}
 \right \}
\ee
where
\be
\fr{\d {\cal G}_{\rm Connected.}}{\d {\breve A}_i} = A^i
\ee
\be
\fr{\d {\cal G}_{\rm Connected.}}{
 {\breve \y^{\a}}_i
} = \y^i_{\a}
\ee
\be
\fr{\d {\cal G}_{\rm 1PI}}{\d {A}^i} =- {\breve A}_i
\ee
\be
\fr{\d {\cal G}_{\rm 1PI}}{\d \y^i_{\a}
} = 
 {\breve \y^{\a}}_i
\ee
and then the identity above in equation (\ref{correctone}) is equivalent to:

\[
\int d^4 x \: \lt \{
\fr{\d {\cal G}_{\rm 1PI} }{\d \G_i} 
\fr{\d {\cal G}_{\rm 1PI} }{\d A^i} 
+
\fr{\d {\cal G}_{\rm 1PI} }{\d {\ov \G}^i} 
\fr{\d {\cal G}_{\rm 1PI} }{\d {\ov A}_i} 
\rt.
\]
\be
\lt.
+
\fr{\d {\cal G}_{\rm 1PI} }{\d {\ov \y}_{i \dot \b}} 
\fr{\d {\cal G}_{\rm 1PI} }{\d {\ov Y}^{i \dot \b} } 
+
\fr{\d {\cal G}_{\rm 1PI} }{\d {  \y}^i_{   \b}} 
\fr{\d {\cal G}_{\rm 1PI} }{\d {  Y}_i^{   \b} } 
\rt \}
+
\fr{\pa {\cal G}_{\rm 1PI} }{\pa { \x}^{\a \dot    \b}} 
\fr{\pa {\cal G}_{\rm 1PI} }{\pa { X}_{\a \dot    \b}} 
=0
\ee
which we will abbreviate to
\be
\la{star1PI}
{\cal G}_{\rm 1PI}* 
{\cal G}_{\rm 1PI}
=0
\ee
Here we can use the loop expansion:
\be
 {\cal G}_{\rm 1PI} =  {\cal A}_{\rm Physical} 
+
 {\cal G}_{\rm 1PI-One \; Loop}
+
 {\cal G}_{\rm 1PI-Two \; Loop}
+\cdots
\la{loopexp}
\ee

Note that we have the following identity from zero loops:
\be
\la{star1PIphys}
{\cal A}_{\rm Physical} * 
{\cal A}_{\rm Physical} =0
\ee

\subsection{Boundary Operator $\d$}

\la{boundaryopinphyspform}

Now we have a new nilpotent operator that is the `square root' of the BRS-ZJ identity:

\[
\d  = 
\int d^4 x \; 
\left \{
\fr{\d {\cal A}_{\rm Physical} }{\d A^i }  
\fr{\d   }{\d \G_i } 
+
\fr{\d {\cal A}_{\rm Physical} }{\d \G_i } 
\fr{\d   }{\d A^i }  
\rt.
\]
\[
\left.
+
\fr{\d {\cal A}_{\rm Physical} }{\d {\ov A}_i } 
\fr{\d  }{\d {\ov \G}^i } 
+
\fr{\d {\cal A}_{\rm Physical} }{\d {\ov \G}^i } 
\fr{\d  }{\d {\ov A}_i } 
\rt.
\]
\[
\left.
+
\fr{\d {\cal A}_{\rm Physical} }{\d \y^{i \a} } 
 \fr{\d   }{\d  Y_{i  \a}}
+
 \fr{\d {\cal A}_{\rm Physical} }{\d  Y_{i  \a}}
\fr{\d   }{\d \y^{i \a} } 
\rt.
\]
\[
\left.
+
\fr{\d {\cal A}_{\rm Physical} }{\d {\ov \y}_{i}^{\dot  \a} }  
\fr{\d  }{\d  {\ov Y}^{i}_{ \dot  \a}}
+
\fr{\d {\cal A}_{\rm Physical} }{\d  {\ov Y}^{i}_{ \dot  \a}}
\fr{\d  }{\d {\ov \y}_{i}^{\dot  \a} }  
\right \}
\]
\be
+
\fr{\pa {\cal A}_{\rm Physical} }{\pa X_{\a \dot \b  }  }  
\fr{\pa  }{\pa  \x^{\a \dot \b  }  } 
+
\fr{\pa {\cal A}_{\rm Physical} }{\pa  \x^{\a \dot \b  }  }  
\fr{\pa }{\pa X_{\a \dot \b  }  }  
\ee 
The  explicit form of this new $\d$ is summarized in 
Table \ref{physicaltable}, which uses composite terms defined in Table \ref{compterms}.

The equation
\be
\d^2 = 0
\ee
follows from equation (\ref{star1PIphys}), as does the equation:
\be
\d {\cal A}_{\rm Physical}
 = 0
\ee
One can also verify these explicitly using Table \ref{physicaltable} and Table \ref{compterms}.

\begin{table}
\caption{   Transformations  for the Physical Formulation of the Massless BRS-ZJ Identity}
\la{physicaltable}
\vspace{.1in}
\framebox{{ $\begin{array}{lll}  
\\
\d A^i&= & 
\fr{\d {\cal A}}{\d \G_i} 
=  \y^{i}_{  \b} {C}^{  \b} 
+ \x^{\g \dot \d} \partial_{\g \dot \d} A^i
\\
\d {\ov A}_i&= & 
\fr{\d {\cal A}}{\d {\ov \G}^i} 
=  {\ov \y}_{i  \dot \b} {\ov C}^{ \dot  \b} 
+ \x^{\g \dot \d} \partial_{\g \dot \d} {\ov A}_i
\\

\d \y_{\a}^i &  =& 
\fr{\d {\cal A}}{\d {  Y}_i^{   \a} } = 
\pa_{ \a \dot \b }  A^{i} {\ov C}^{\dot \b}  
+ 
C_{\a}   
G^i
+ \x^{\g \dot \d} \partial_{\g \dot \d}  \y^{i}_{\a  }
\\

\d
 {\ov \y}_{i \dot \a} &  =& 
\fr{\d {\cal A}}{\d { {\ov Y}}^{i \dot   \a} } = 
\pa_{ \a \dot \a }  {\ov A}_{i} {C}^{\a}  
+ 
{\ov C}_{\dot \a}   
{\ov G}_{i}
+ \x^{\g \dot \d} \partial_{\g \dot \d} 
 {\ov \y}_{i \dot \a} 
\\
\d \G_i 
&= &
 \fr{\d {\cal A}}{\d A^i} 
=
 - \fr{1}{2} \pa_{ \a \dot \b  }       \pa^{ \a \dot \b  }        {\ov  A}_{i} 
 + g_{ijk} { G}^{jk}
\\
&&
-\pa_{ \a \dot \b } Y_{i}^{ \a}    {\ov C}^{\dot \b}   
+ \x^{\g \dot \d} \partial_{\g \dot \d} \G_i
\\

\d {\ov \G}^i 
&= & \fr{\d {\cal A}}{\d {\ov A}_i} 
=
- \fr{1}{2} \pa_{ \a \dot \b  }       \pa^{ \a \dot \b  }        { A}^{i} 
+  {\ov g}^{ijk}     {\ov  G}_{j k} 
\\
&&
- \pa_{ \a \dot \b } {\ov Y}^{ i \dot \b}    {C}^{\a}   
+ \x^{\g \dot \d} \partial_{\g \dot \d} 
 {\ov \G}^i
\\
\d Y_{i}^{ \a} 
&=&\fr{\d {\cal A}}{\d {  \y}^i_{   \a}} 
= 
-
  \pa^{\a \dot \b  }   
{\ov \y}_{i   \dot \b}
 +
2 g_{ijk}  \y^{j \a} A^k    
-
\G_i  
 {C}^{  \a}
+ \x^{\g \dot \d} \partial_{\g \dot \d}  Y_{i}^{ \a}
\\
\d 
{\ov Y}^{i \dot \a} 
&=&\fr{\d {\cal A}}{\d {\ov \y}_i^{ \dot \a} 
} 
= 
-
  \pa^{\b \dot \a  }   
{ \y}^i_{ \b}
\\
&&
+
2 {\ov g}^{ijk} {\ov \y}_{j}^{\dot  \a} 
{\ov A}_k  
-
{\ov \G}^i  
 {\ov C}^{\dot  \a}
+ \x^{\g \dot \d} \partial_{\g \dot \d}  
{\ov Y}^{i \dot \a} 
\\
 \d G^i 
&=&
  \pa_{\a \dot \b}   \y^{i \a} {\ov C}^{\dot \b} 
+ \x^{\g \dot \d} \partial_{\g \dot \d}  G^i 
\\
 \d G^{ij} &=& \pa_{\a \dot \b} \lt ( A^i  \y^{j \a} + A^j  \y^{i \a} \rt )  {\ov C}^{\dot \b} 
\\
 \d G^{ijk} &=& \pa^{\a \dot \b} \lt ( A^{i}   A^{j} \y^{k}_{\a}  + A^{j}   A^{k} \y^{i }_{\a}  + A^{k}   A^{i} \y^{j}_{\a} \rt )  {\ov C}_{\dot \b} 
\\
\d \x_{\a \dot \b} 
&=& \fr{\pa {\cal A}}{\pa { X}^{\a \dot    \b}} 
=
 -   C_{\a} {\ov C}_{\dot \b}
\\
\d X_{\a \dot \b} 
&=& \fr{\pa {\cal A}}{\pa { \x}^{\a \dot    \b}} 
= 
\int d^4 x \; \X_{\a \dot \b} 
\\
\d C_{\a}
&=&
0
\\
\d  {\ov C}_{\dot \b}
&=&
0

\end{array}$}} 
\end{table}

\begin{table}
\caption{  Composite Terms $G$  for Massless   Chiral Supersymmetry}
\la{compterms}
\vspace{.1in}
\framebox{{ $\begin{array}{lll}  
\\
 G^i 
&=&
- \lt ( 
 {\ov g}^{ijk} {\ov A}_j  {\ov A}_k +
{\ov Y}^{i \dot \b} {\ov C}_{\dot \b } 
\rt )
\\
G^{ij}&= & 
     A^{i}   
G^j
 +     A^{j}   
G^i
- \y^{i \a} 
\y^{j}_{ \a}  
 \\
G^{(ijk)}
&= & 
 A^{i}  A^{j}  G^k
+ A^{j}  A^{k}  G^i
+ A^{k}  A^{i}  G^j
\\&&
 - 
 \y^{i \a} \y^{j}_{ \a}  A^k
 - 
 \y^{j \a} \y^{k}_{ \a}  A^i
 - 
 \y^{k \a} \y^{i}_{ \a}  A^j
\\
\\
{\ov G}_{i} 
&=&
-\lt ( 
{g}_{ilq} A^l A^q
+
{Y}_i^{  \b} {C}_{\b } 
\rt )
\\
{\ov G}_{ij} 
&=&
\A_i \ov G_{j} 
+\A_j \ov G_{i} 
-
\oy_i^{\dot \b}  
\oy_{j,\dot \b}  
\\
&=&
-\A_j \lt (  
{g}_{ilq} A^l A^q 
+
{Y}_i^{  \b} {C}_{\b } 
\rt )
\\
&&
-
\A_i \lt ( 
{g}_{jlq} A^l A^q
+
{Y}_j^{  \b} {C}_{\b } 
\rt )
-
\oy_i^{\dot \b}  
\oy_{j,\dot \b}  
\\
\\
\X_{\g \dot \d}
&=&
 Y_{i}^{ \a}  \pa_{\g \dot \d}  \y^{i}_{ \a}
+
 {\ov Y}^{i \dot \b}  \pa_{\g \dot \d}{\ov \y}_{i \dot \b}  
-
 \G_{i}   \pa_{\g \dot \d}  A^{i} 
-
 {\ov \G}^{i  }  \pa_{\g \dot \d}{\ov A}_{i }  
\\

\end{array}$}} 
\end{table}

\subsection{Derivative Form of $\d$ }
\la{formofdeltaWZ1}
Another way to write $\d$  for the massless interacting case is:
\be
\d =
\int d^4 x\;
\lt (
  \y^{i}_{  \b} {C}^{  \b} 
+ \x^{\g \dot \d} \partial_{\g \dot \d} A^i
\rt )
\fr{\d  }{\d A^i} 
\ee

\be
+
\int d^4 x\;
\lt (
\pa_{ \a \dot \b }  A^{i} {\ov C}^{\dot \b}  
+ 
C_{\a}   
G^i
+ \x^{\g \dot \d} \partial_{\g \dot \d}  \y^{i}_{\a  }
\rt )
\fr{\d  }{\d \y_{\a}^i} 
\ee

\be
+
\int d^4 x\;
\lt (
 - \fr{1}{2} \pa_{ \a \dot \b  }       \pa^{ \a \dot \b  }        {\ov  A}_{i} 
   + g_{ijk} G^{jk}
-
\pa_{ \a \dot \b } Y_{i}^{ \a}    {\ov C}^{\dot \b}   
+ \x^{\g \dot \d} \partial_{\g \dot \d} \G_i
\rt )
 \fr{\d}{\d \G_i } 
\ee

\be
+
\int d^4 x\;
\lt (
-
  \pa^{\a \dot \b  }   
{\ov \y}_{i   \dot \b}
 +
2 g_{ijk}  \y^{j \a} A^k    
-
\G_i  
 {C}^{  \a}
+ \x^{\g \dot \d} \partial_{\g \dot \d}  Y_{i}^{ \a}
\rt )
 \fr{\d  }{\d Y_{i}^{ \a}} 
\ee
\be
+  {\rm Complex\; Conjugate}
+
\int d^4 x \; \X_{\a \dot \b} 
 \fr{\pa }{\pa { X}^{\a \dot    \b}} 
 -   C_{\a} {\ov C}_{\dot \b}
\fr{\pa  }{\pa { \x}^{\a \dot    \b}} 
\ee

where 
\be
 G^i 
- \lt ( 
 {\ov g}^{ijk} {\ov A}_j  {\ov A}_k +
{\ov Y}^{i \dot \b} {\ov C}_{\dot \b } 
\rt )
\ee
and
\be
G^{ij}=
     A^{i}   
G^j
 +     A^{j}   
G^i
- \y^{i \a} 
\y^{j}_{ \a}  
\ee

We will drop the term
\be
\int d^4 x \; \X_{\a \dot \b} 
 \fr{\pa }{\pa { X}^{\a \dot    \b}} 
\la{fqwefwewef}
\ee for now.  The resulting $\d$ is still nilpotent, and dropping this term eliminates one part from the cohomology that seems to have little importance at this stage.

\subsection{Expanded Form of $\d$}
\la{formofdeltaWZ2}

If we expand everything explicitly, and drop the term 
(\ref{fqwefwewef}), this becomes:
\be
\d =
\int d^4 x\;
\lt (
  \y^{i}_{  \b} {C}^{  \b} 
+ \x^{\g \dot \d} \partial_{\g \dot \d} A^i
\rt )
\fr{\d  }{\d A^i} 
\ee

\be
+
\int d^4 x\;
\lt (
\pa_{ \a \dot \b }  A^{i} {\ov C}^{\dot \b}  
- 
C_{\a}   
 \lt ( 
 {\ov g}^{ijk} {\ov A}_j  {\ov A}_k +
{\ov Y}^{i \dot \b} {\ov C}_{\dot \b } 
\rt )
+ \x^{\g \dot \d} \partial_{\g \dot \d}  \y^{i}_{\a  }
\rt )
\fr{\d  }{\d \y_{\a}^i} 
\ee

\be
+
\int d^4 x\;
\lt (
 - \fr{1}{2} \pa_{ \a \dot \b  }       \pa^{ \a \dot \b  }        {\ov  A}_{i} 
\ebp
   + g_{ijk} \lt [
-     A^{j}   
\lt ( 
 {\ov g}^{klm} {\ov A}_l  {\ov A}_m +
{\ov Y}^{k \dot \b} {\ov C}_{\dot \b } 
\rt )
 -     A^{k}   
 \lt ( 
 {\ov g}^{jlm} {\ov A}_l  {\ov A}_m +
{\ov Y}^{j \dot \b} {\ov C}_{\dot \b } 
\rt )
- \y^{j \a} 
\y^{k}_{ \a}  
 \rt ]
\ebp
-
\pa_{ \a \dot \b } Y_{i}^{ \a}    {\ov C}^{\dot \b}   
+ \x^{\g \dot \d} \partial_{\g \dot \d} \G_i
\rt )
 \fr{\d}{\d \G_i } 
\ee

\be
+
\int d^4 x\;
\lt (
-
  \pa^{\a \dot \b  }   
{\ov \y}_{i   \dot \b}
 +
2 g_{ijk}  \y^{j \a} A^k    
-
\G_i  
 {C}^{  \a}
+ \x^{\g \dot \d} \partial_{\g \dot \d}  Y_{i}^{ \a}
\rt )
 \fr{\d  }{\d Y_{i}^{ \a}} 
\ee
\be
+ {\rm Complex \; Conjugate}
 -   C_{\a} {\ov C}_{\dot \b}
\fr{\pa  }{\pa { \x}^{\a \dot    \b}} 
\ee

\section{Simple Dotspinors and Undotspinors: General Description of Simple Generators}
\la{fdaffsdfa}

In \ci{cybersusyI} we wrote down the first parts of certain expressions for composite operators with the quantum numbers of the electron and positron, and we claimed there that they could be built up into composite chiral dotted spinor superfields. 

In fact, those leptonic dotspinors originate in the cohomology space of the operator $\d$ in section \ref{formofdeltaWZ2}. However, they refer to the SSM and we must wait for \ci{cybersusyIII} to discuss them.  Here we prepare for that in a general way.

First let us explain how to generate the simple dotspinors and undotspinors for the general action and $\d$ discussed above.  
\ben
\item
Firstly we have the simple  generators $\w_{(\dot \a_1 \cdots \dot \a_{2m+1})}$. 	  These have the form:
\be
 \w_{(\dot \a_1 \cdots \dot \a_{2m+1})} = f_{(j_1 \cdots j_n)}^{[i_1 \cdots i_{2m+1}]}
\oy_{i_1 \dot \a_1}  \cdots \oy_{i_{2m+1} \dot \a_{2m+1}}   A^{j_1} \cdots A^{j_n}; 
\la{dotpspinorgeneralfermion}
\eb
 \; m=0, 1,2 \cdots ,n= 0,1,2 \cdots 
\ee
This expression has a number of symmetry properties:\ben
\item
Note that $ \w_{\dot \a_1 \cdots \dot \a_{2m+1}}$ has an odd number of dotspinor indices, so that it is a fermion. We indicate this by using a Greek letter to describe it, which is a general convention used here.  Latin letters describe bosons.
\item
The brackets $(\cdots )$ around the dotspinor indices $(\dot \a_1 \cdots \dot \a_{2m+1})$ 
indicate that these indices are symmetrized. This means that the spin of $  \w_{\dot \a_1 \cdots \dot \a_{2m+1}}$ is $J= \fr{2m+1}{2}$, which is consistent with it being a fermion.
\item
The brackets $(\cdots )$ around the flavour indices $(j_1 \cdots j_n)$ 
indicate that these indices are symmetrized, which is automatic because the fields $ A^{j_1} $ are commuting quantities.
\item
The brackets $[\cdots ]$ around the flavor indices $[i_1 \cdots i_{2m+1}]$ 
indicate that these indices are antisymmetrized.  This antisymmetrization is mandated by the symmetry on $(\dot \a_1 \cdots \dot \a_{2m+1})$  and the fact that the spinors $\oy_{i_1\dot \a_1}  $ are anticommuting. 
\item
In addition,  the dimensionless numerical coefficients $f_{j_1 \cdots j_n}^{i_1 \cdots i_{2m+1}}$ obeys a constraint, which arises from the following equation: 
\be
 d_3 \ov \w_{(\dot \a_1 \cdots \dot \a_{2m+1})}
=0
\ee
which, in detail, is:
\be
 \lt \{
\oC_{\dot \a} {g}_{ijk} A^j A^k 
\oy_{i \dot \a}^{ \dag}\rt \} 
f_{(j_1 \cdots j_n)}^{[i_1 \cdots i_{2m+1}]}
\oy_{i_1 \dot \a_1}  \cdots \oy_{i_{2m+1} \dot \a_{2m+1}}   A^{j_1} \cdots A^{j_n}
=0
\la{ffweweffwe}
\ee
This can be written as a constraint on the coefficients by eliminating the fields by differentiation:
\be
 {g}_{s (j_{n+1} j_{n+2}}   
f_{ j_1 \cdots j_{n})}^{[s i_2 \cdots i_{2m+1}]}
=0
\la{ffweweffwe22}
\ee
where the expression must be symmetrized over the indices $(j_1 \cdots j_{n} j_{n+1} j_{n+2} )$
\item
The notation 
\be
 d_3  
= 
\oC_{\dot \a} {g}_{ijk} A^j A^k 
\oy_{i \dot \a}^{ \dag} 
+{\rm Complex\; Conjugate}
\ee
comes from the spectral sequence analysis, which was introduced in \ci{bigpaper}.  We do not use that analysis here since it is too incomplete to describe yet.  All the results here are proved explicitly instead, which is really more useful for present purposes anyway. We shall see that this operator $d_3$ has a natural explanation in terms of the full theory, as appears in section \ref{qfwqergreg} below.

\een
\item
With suitable  changes, all the above remarks also apply to the simple generators $\ov \w_{(\a_1 \cdots \a_{2m+1})} $, which are  the complex conjugates of 
(\ref{dotpspinorgeneralfermion}).    These have the form: 
\be
\ov \w_{(\a_1 \cdots \a_{2m+1})} = f^{(j_1 \cdots j_n)}_{[i_1 \cdots i_{2m+1}]}
\y^{i_1}_{(\a_1}  \cdots \y^{i_{2m+1}}_{\a_{2m+1})}   \A_{j_1} \cdots \A_{j_n}; \la{undotpspinorgeneralfermion}
\eb
 \; m=0, 1,2 \cdots ,n= 0,1,2 \cdots 
\la{fwweffwefwefw}
\ee
The complex conjugate constraint is:
\be
 \lt \{
C_{ \a} {\ov g}^{ijk} \A_j \A_k 
\y_{ \dot \a}^{i \dag}\rt \} 
f^{(j_1 \cdots j_n)}_{[i_1 \cdots i_{2m+1}]}
\y^{i_1}_{(\a_1}  \cdots \y^{i_{2m+1}}_{\a_{2m+1})}   \A_{j_1} \cdots \A_{j_n};  
\la{undotpspinorgeneralfermion}\eb
 \; m=0, 1,2 \cdots ,n= 0,1,2 \cdots 
\ee
\item
With suitable changes, these remarks also apply to the   simple generators  $A_{(\dot \a_1 \cdots \dot \a_{2m})} $.  These are bosons because they have integer spin 
$J = \fr{2m}{2} = m$.  These have the form:

\be
 A_{(\dot \a_1 \cdots \dot \a_{2m})} = f_{(j_1 \cdots j_n)}^{[i_1 \cdots i_{2m}]}
\oy_{i_1 (\dot \a_1}  \cdots \oy_{i_{2m} \dot \a_{2m})}   A^{j_1} \cdots A^{j_n}; \la{dotpspinorgeneralboson}
\eb
 \; m=0, 1,2 \cdots ,n= 0,1,2 \cdots 
\ee
\item
With suitable changes, these remarks also apply to the     simple generators  $\ov A_{(\a_1 \cdots \a_{2m})}$.  These are   bosons because they have integer spin 
$J = \fr{2m}{2} = m$.     These have the form:
\be
\ov A_{(\a_1 \cdots \a_{2m})} =
f^{(j_1 \cdots j_n)}_{[i_1 \cdots i_{2m}]}\y^{[i_1}_{(\a_1}  \cdots \y^{i_{2m}]}_{\a_{2m})}   \A_{(j_1} \cdots \A_{j_n)}; 
\la{undotpspinorgeneralboson}
\eb
 \; m= 1,2 \cdots ,n= 0,1,2 \cdots 
\ee

\een

In the next sections we shall explain how to build these simple generators into full expressions which are in the cohomology space of the operator $\d$ defined by Table 
\ref{physicaltable}.  To do that we will review superspace first in our notation, and then a new kind of construction which we call pseudosuperspace. Pseudosuperspace is just superspace, but the components of the superfields are composite, so we have to be a bit careful.

\section{Quick Review of Superspace}

\subsection{Superspace Notation}

\la{wefqwhjhuynfbn}

$D_{\a}$ is the superspace covariant derivative, defined by
\be
D_{\a} = \fr{\pa}{\pa \q^{\a}} + 
\fr{1}{2}\ov \q^{\dot \b} \pa_{\a \dot \b}
\ee
The complex conjugate superspace covariant derivative  is  defined by:
$\ov D_{\dot \a}$
\be
\ov D_{\dot \a} = \fr{\pa}{\pa \ov \q^{\dot \a}} + 
\fr{1}{2}\q^{\b} \pa_{\b \dot \a}
\ee
We also define the superspace translation generator by
$Q_{\a}$  
\be
Q_{\a} = \fr{\pa}{\pa \q^{\a}} - 
\fr{1}{2}\ov \q^{\dot \b} \pa_{\a \dot \b}
\ee
and  complex conjugate superspace translation generator   is  defined by:
$\ov Q_{\dot \a}$
\be
\ov Q_{\dot \a} = \fr{\pa}{\pa \ov \q^{\dot \a}} - 
\fr{1}{2}\q^{\b} \pa_{\b \dot \a}
\ee
We also define the chirally translated quantity
\be
 y_{\g \dot \d}= x_{\g \dot \d}
+ \fr{1}{2} \q_{\g} {\ov \q}_{ \dot \d}
\ee
where $x_{\g \dot \d}$ are the coordinates of spacetime. It satisfies:
\be
{\ov D}_{\dot \a} y_{\g \dot \d}=0
\ee

\subsection{Superspace Expansion}

\la{wefqwhjhuynfbn2}

First we shall recall some standard superfield theory, using our notation. 

\subsubsection{Chiral Scalar Superfields}

${\widehat A}^{p}$ is used to describe an arbitrary  set of chiral scalar superfields and $p=1...n$  is an index to distinguish among the members of the set.  
An expansion in superspace can be written in the compact form:
\be
{\widehat A}^{p}(x)= A^{p}(y) +   \q^{\a} \y^{p}_{ \a}(y)+ 
\fr{1}{2} \q \cdot \q F^{p}(y)
\ee
where $A^{p}$ is a set of scalar fields, $\y^{p}_{\a}$ is a set of spinor fields (where $\a=1,2$ is a two-component Weyl spinor index), and $F^{p}$ are a set of auxiliary scalar fields.

This satisfies the constraint
\be
{\ov D}_{\dot \a} {\widehat A}^{p}(x)=0
\ee
which is satisfied so long as the parameter satisfies 
\be
{\ov D}_{\dot \a} y_{\g \dot \d}=0
\ee 
which means that 
\be
 y_{\g \dot \d}= x_{\g \dot \d}
+ \fr{1}{2} \q_{\g} {\ov \q}_{ \dot \d}
\ee
where $x_{\g \dot \d}$ are the coordinates of spacetime.

The components transform in the following way:
\be
\vspace{.1in}
 \begin{array}{lll}  
\\
\d A^p&= & 
 \y^p_{\a} {C}^{  \a} 
+ \x^{\g \dot \d} \partial_{\g \dot \d} A^p 
\\
 \d { \y}^p_{\a} &  =& 
\pa_{ \a \dot \b }  A^p \ov C^{\dot \b} + F^p  { C}_{ \a}  
+ \x^{\g \dot \d} \partial_{\g \dot \d}   { \y}_{\a}^p 
\\
\d
 F^p  &  =&  \pa_{ \b \dot \g }  \y^{p \b}  \ov C^{\dot \g} 
 + \x^{\g \dot \d} \partial_{\g \dot \d}  F^p
\\
\end{array}
\ee

When we use a different letter than ${\widehat A}$ to indicate the superfield, for example $\widehat B$, we will use a notation that indicates this by writing $B$,  $\y_B$ and $F_B$ for the other components. 

Thus we have:
\be
{\widehat A}^{p}(x)= A^{p}(y) +   \q^{\a} \y^{p}_{ \a}(y)+ 
\fr{1}{2} \q \cdot \q F^{p}(y)
\ee
but
\be
{\widehat B}^{p}(x)= B^{p}(y) +   \q^{\a} \y^{p}_{B, \a}(y)+ 
\fr{1}{2} \q \cdot \q F^{p}_{B}(y)
\la{grthhrtjhrt}
\ee
This is a useful notation when we come to the SSM, since there are many different superfields there.

\subsubsection{Chiral Dotted  Spinor Superfields}

${\widehat \w}_{p \dot \a}$ is used to describe an arbitrary  set of dotted chiral spinor superfields and $p=1...n$  is an index to distinguish among the members of the set.  An expansion in superspace can be written in the compact form:
\be
{\widehat \w}_{p \dot \a}(x)= \w_{p  \dot \a}(y) +  \q^{\d} W_{p \d \dot \a}(y)+ 
\fr{1}{2} \q \cdot \q \L_{p \dot \a}(y)
\ee
where $\w_{p  \dot \a}$ is a set of dotted spinor fields, $W_{p \d \dot \a}$ is a set of vector fields, and $\L_{p \dot \a}$ are a set of dotted spinor fields.
This satisfies the constraint
\be
{\ov D}_{\dot \a} {\widehat \w}_{p \dot \b}=0
\ee

It transforms in the following way:
\be
\vspace{.1in}
 \begin{array}{lll}  
\\
\d {  {\w}}_{ p \dot \a}&= & 
  { W}_{p \a \dot   \a} {C}^{  \a} 
+ \x^{\g \dot \d} \partial_{\g \dot \d} { \w}_{p \dot \a}
\\
 \d { W}_{p\a \dot \a} &  =& 
\pa_{ \a \dot \b }  { \w}_{p \dot \a} \ov C^{\dot \b} + { \L}_{p \dot \a} {\ C}_{ \a}  
+ \x^{\g \dot \d} \partial_{\g \dot \d}   {  W}_{p \a \dot \a} 
\\
\d
 { \L}_{p \dot \a} &  =&  \pa_{ \b \dot \g }  { W}_{p \;\;\dot \a}^{ \b}  \ov C^{\dot \g} 
 + \x^{\g \dot \d} \partial_{\g \dot \d}  { \L}_{p \dot \a}
\\
\end{array}
\la{fqwertgreger}
\ee
This is just the transformation of a chiral scalar, with an extra index $\dot \a$ carried along inertly.

The complex conjugate is:
\be
{\widehat {\ov \w}}^i_{  \a}(x)= \ov \w^i_{   \a}(\ov y) +  \ov \q^{\dot \d} \ov W^i_{ \d \dot \a}(\ov y)+ 
\fr{1}{2} \ov \q \cdot \ov \q \ov \L^i_{  \a}(\ov y)
\ee
It transforms in the following way:
\be
\vspace{.1in}
 \begin{array}{lll}  
\\
\d {\hat {\ov \w}}^i_{  \a}&= & 
  {\ov W}^i_{\a \dot   \b} {\oC}^{\dot   \b} 
+ \x^{\g \dot \d} \partial_{\g \dot \d} {\ov \w}^i_{\a}
\\
 \d {\ov W}^i_{\a \dot \b} &  =& 
\pa_{ \g \dot \b }  {\ov \w}^i_{\a} C^{\g} + {\ov \L}^i_{\a} {\ov C}_{\dot \b}  
+ \x^{\g \dot \d} \partial_{\g \dot \d}   {\ov W}^i_{\a \dot \b} 
\\
\d
 {\ov \L}^i_{\a} &  =&  \pa_{ \g \dot \b }  {\ov W}_{\a}^{i\;\; \dot \b}  C^{\g} 
 + \x^{\g \dot \d} \partial_{\g \dot \d}  {\ov \L}^i_{\a}
\\
\end{array}
\ee

When we use a different letter to indicate the superfield, for example ${\widehat \f}_{p \dot \a}$, so that its spinorial component is $\f_{p  \dot \a}$, we will use a notation that indicates this by writing $W_{\f,p \d \dot \a}$ and $\L_{\f,p \dot \a}$ for the other components. This is similar to (\ref{grthhrtjhrt}).
Thus for example:

\be
{\widehat \f}_{p \dot \a}(x)= \f_{p  \dot \a}(y) +  \q^{\d} W_{\f,p \d \dot \a}(y)+ 
\fr{1}{2} \q \cdot \q \L_{\f,p \dot \a}(y)
\la{wfwefwefjkljklljk}
\ee

\section{Pseudosuperspace}

Since we have integrated the auxiliary in defining our operator $\d$, we no longer can use superspace.  However there is a replacement for superspace which works in a similar way.  The differences are important however.

\subsection{The fundamental chiral  composite scalar pseudosuperfield}

First we define the `fundamental chiral composite scalar pseudosuperfield':
\be
{\hat A}^{i}(x) = A^i(y)
+
\q^{\b} 
\y^i_{\b}(y)
+ \fr{1}{2} \q^{\g} \q_{\g} 
G^i(y)
\la{qgregerjtjt}
\ee
where $G^i$ is defined in Table
 \ref{compterms}.
This is a  chiral pseudosuperfields and it satisfies:
\be
\ov D_{\dot \d}{\hat A}^i(x)
=0
\ee
It is easy to verify that ${\hat A}_{i}$ really does transform, under the action of $\d$ defined by (\ref{physicaltable}),
just like an ordinary superfield, so we can write:
\be
\d {\hat A}^{i} =  ( C\cdot Q +\ov C\cdot \ov Q ) {\hat A}^{i} 
\la{fwefwfwef2}
\ee
In other words ${\hat A}^{i}$ transforms like the fields in 
equation (\ref{Purechiral1}), except that the composite field
 $G^i$ has taken the place of $F^i$ in those equations. 
In other words we have
\be
\vspace{.1in}
 \begin{array}{lll}  
\\
\d A^i&= & 
 \y^i_{\a} {C}^{  \a} 
+ \x^{\g \dot \d} \partial_{\g \dot \d} A^i 
\\
 \d { \y}^i_{\a} &  =& 
\pa_{ \a \dot \b }  A^i \ov C^{\dot \b} + G^i  { C}_{ \a}  
+ \x^{\g \dot \d} \partial_{\g \dot \d}   { \y}_{\a}^i 
\\
\d
 G^i  &  =&  \pa_{ \b \dot \g }  \y^{p \b}  \ov C^{\dot \g} 
 + \x^{\g \dot \d} \partial_{\g \dot \d}  G^i
\\
\end{array}
\ee

\subsection{
The fundamental chiral composite dotted  spinor pseudosuperfields}

Next we define the `fundamental chiral composite  dotted  spinor pseudosuperfield':
\be
{\hat \f}_{i \dot \a}(x) = \ov \y_{\dot \a i}(y) +
\q^{\b} 
\lt (
\pa_{\b \dot \a} \A_i(y) 
+ \ov C_{\dot \a} Y_{i  \b}(y) \rt )
- \fr{1}{2} \q^{\g} \q_{\g} 
\G_{i}(y)  
\ov C_{\dot \a} 
\la{weird}
\ee
From equation (\ref{wfwefwefjkljklljk}), the standard notation for a   chiral   dotted  spinor  superfield
is
\be
{\widehat \f}_{i \dot \a}(x)= \f_{i  \dot \a}(y) +  \q^{\d} W_{\f,i \d \dot \a}(y)+ 
\fr{1}{2} \q \cdot \q \L_{\f,i \dot \a}(y)
\ee
So we make the identifications:
\be
\f_{i  \dot \a}=\ov \y_{\dot \a i}
\la{fund1}
\ee
\be
W_{\f,i \d \dot \a}=
\lt (
\pa_{\b \dot \a} \A_i 
+ \ov C_{\dot \a} Y_{i  \b}\rt )
\la{fund2}
\ee
\be
\L_{\f,i \dot \a}
=-\G_{i}  
\ov C_{\dot \a} 
\la{fund3}
\ee

This is a chiral pseudosuperfield:
\be
\ov D_{\dot \d}{\hat \f}_{i \dot \a}(x) 
=0
\ee
It turns out that the transformations   $\d$ defined by 
(\ref{formofdeltaWZ2}) induce transformations on this composite expression ${\hat \f}_{i \dot \a}$ so that it transforms  as a dotted chiral spinor superfield, except that there are extra terms in addition to the usual ones (so it is not really a superfield at all, in the usual sense):
\be
\d {\hat \f}_{i \dot \a} =  ( C\cdot Q +\ov C\cdot \ov Q ) {\hat \f}_{i \dot \a}   -
 g_{ijk} 
{\hat A}^j {\hat A}^k \ov C_{\dot \a}
+ \x^{\g \dot \d} \partial_{\g \dot \d} {\hat \f}_{i \dot \a}
\la{fwefwfwef}
\ee
This means the following in components: 
\be
\vspace{.1in}
 \begin{array}{lll}  
\\
\d {  {\f}}_{ i \dot \a}&= & 
  { W}_{\f,i \a \dot   \a} {C}^{  \a} 
-
 g_{ijk} 
A^j A ^k \ov C_{\dot \a}
+ \x^{\g \dot \d} \partial_{\g \dot \d} { \f}_{i \dot \a}
\\
 \d { W}_{\f, i\a \dot \a} &  =& 
\pa_{ \a \dot \b }  { \f}_{i \dot \a} \ov C^{\dot \b} + { \L}_{i \dot \a} { C}_{ \a}  
+
2 g_{ijk} 
A^j \y^k_{\a} \ov C_{\dot \a}
+ \x^{\g \dot \d} \partial_{\g \dot \d}   {  W}_{\f, i \a \dot \a} 
\\
\d
 { \L}_{\f, i \dot \a} &  =&  \pa_{ \b \dot \g }  { W}_{\f, i \;\;\dot \a}^{ \b}  \ov C^{\dot \g} 
-
 g_{ijk} 
\lt (2 A^j G^k -\y^{j \b}\y^{k}_{ \b}\rt )
\ov C_{\dot \a}
 + \x^{\g \dot \d} \partial_{\g \dot \d}  { \L}_{\f,i \dot \a}
\\
\end{array}
\la{wetqtgtgerhhrt}
\ee

We shall demonstrate these transformations for the complex conjugate, which follows in the next section.

\section{
The fundamental antichiral composite undotted  spinor pseudosuperfields}

The complex conjugate of (\ref{weird}) 
is:
\be
{\hat {\ov \f}}^i_{\a}(x) =\y^i_{\a }(\ov y) +
\ov \q^{\dot \b} 
\lt (
\pa_{\a \dot \b} A^i(\ov y) 
+  C_{\a} \ov Y^i_{\dot  \b}(\ov y) \rt )
- \fr{1}{2} \ov \q^{\dot \g} \ov \q_{\dot \g} 
\ov \G^{i}(\ov y)  
C_{ \a} 
\la{weirdcc}
\ee
and the standard notation is:
\be
{\widehat {\ov \f}}^i_{  \a}(x)= \ov \f^i_{   \a}(\ov y) +  \ov \q^{\dot \d} \ov W^i_{\f, \d \dot \a}(\ov y)+ 
\fr{1}{2} \ov \q \cdot \ov \q \ov \L^i_{\f,  \a}(\ov y)
\ee
So
we identify
\be
{\ov \f}^i_{\a  }   = 
\y^i_{\a } 
\la{fund1cc}
 \ee
\be
{\ov W}^i_{\f,\a \dot \b}   =
\lt (
\pa_{\a \dot \b} A^i 
+  C_{\a} \ov Y^i_{\dot  \b}  \rt )
\la{fund2cc}
\ee
\be
{\ov \L}^i_{\f,\a }   =
-\ov \G^{i}   
C_{ \a} 
\la{fund3cc}
\ee

We will now show that this composite expression ${\ov {\hat \f}}^i_{ \a}$ does transform as an undotted antichiral spinor superfield, except that there are extra terms in addition to the usual ones, as follows:
\be
\d {\hat {\ov \f}}^i_{ \a} =  ( C\cdot Q +\ov C\cdot \ov Q )
 {\hat {\ov \f}}^i_{ \a}   -
 \ov g^{ijk} 
{\hat {\ov A}}_j {\hat {\ov A}}_k  C_{\a}
+ \x^{\g \dot \d} \partial_{\g \dot \d}{\hat {\ov \f}}^i_{ \a}
\la{fwefwfwefcc}
\ee
We can ignore the $\x^{\g \dot \d} \partial_{\g \dot \d}$ terms. In components this is :
\be
\vspace{.1in}
 \begin{array}{lll}  
\\
\d {  {\ov \f}}^i_{  \a}&= & 
  {\ov W}^i_{\f,\a \dot   \b} {\oC}^{\dot   \b} 
- 
C_{\a}   
 {\ov g}^{ijk} {\ov A}_j  {\ov A}_k 
\\
 \d {\ov W}^i_{\f, \a \dot \b} &  =& 
\pa_{ \g \dot \b }  {\ov \f}^i_{\a} C^{\g} + {\ov \L}^i_{\f, \a} {\ov C}_{\dot \b}  
+ 
2 C_{\a}   
 {\ov g}^{ijk} {\ov A}_j  {\ov \y}_{k \dot \b} 
\\
\d
 {\ov \L}^i_{\f, \a} &  =&  \pa_{ \g \dot \b }  {\ov W}_{\f, \a}^{i\;\; \dot \b}  C^{\g} 
- 
 C_{\a}   
 {\ov g}^{ijk}
 {\ov G}_{jk}
\\
\end{array}
\la{wetqtgtgerhhrtov}
\ee
where we use
\be
{\ov G}_{ij} 
=
\A_i \ov G_{j} 
+\A_j \ov G_{i} 
-
\oy_i^{\dot \b}  
\oy_{j,\dot \b}  
\ee

It is straightforward to verify that the expressions
(\ref{fund1cc}), (\ref{fund2cc}) and (\ref{fund3cc}) do generate these transformations (\ref{wetqtgtgerhhrtov})
 when we use $\d$ defined by 
 Table 
\ref{physicaltable}.  We will now demonstrate this very important fact explicitly:
\subsection{Verification of the first transformation}

The first transformation is:
\be
\d \ov \f^i_{ \a  } =  
 \d \y_{ \a}^{ i}  =   
 \lt (
\pa_{ \a \dot \b }  A^{i} {\ov C}^{\dot \b}  
- 
C_{\a}   
 \lt ( 
 {\ov g}^{ijk} {\ov A}_j  {\ov A}_k +
{\ov Y}^{i \dot \b} {\ov C}_{\dot \b } 
\rt )
\rt )
\eb
=  
 \lt (
\pa_{ \a \dot \b }  A^{i} {\ov C}^{\dot \b}  
+
C_{\a}   
 {\ov Y}^{i}_{ \dot \b}
\rt )  {\ov C}^{\dot \b }  - 
C_{\a}   
 {\ov g}^{ijk} {\ov A}_j  {\ov A}_k 
\eb
=  
\ov W_{\f, \a \dot \b }   {\ov C}^{\dot \b } - 
C_{\a}   
 {\ov g}^{ijk} {\ov A}_j  {\ov A}_k 
 \ee
So this first term is indeed transforming as though it were the lowest term of an antichiral undotted spinor superfield, provided we add on the extra terms $ - 
C_{\a}   
 {\ov g}^{ijk} {\ov A}_j  {\ov A}_k $.

\subsection{Verification of the second transformation}

The next transformation, from (\ref{fund2cc}), is:

\be
\d {\ov W}^i_{\f,\a \dot \b}   =
\d \lt (
\pa_{\a \dot \b} A^i 
+  C_{\a} \ov Y^i_{\dot  \b}  \rt )
\ee
and using  $\d$ defined by 
 Table 
\ref{physicaltable}, we get
\be
\d {\ov W}^i_{\f,\a \dot \b}   
=
\pa_{\a \dot \b}  \y^{i}_{  \g} {C}^{  \g} 
+  C_{\a}
\lt (
  \pa_{\b \dot \b  }   
{ \y}^{i  \b}
\\
+
2 {\ov g}^{ijk} {\ov \y}_{j \dot  \b} 
{\ov A}_k  
-
{\ov \G}^i  
 {\ov C}_{\dot  \b}
\rt )
\eb
=
\pa_{\a \dot \b}  \y^{i}_{  \g} {C}^{  \g} 
+  C_{\a}
  \pa_{\b \dot \b  }   
{ \y}^{i  \b}
+C_{\a}
2 {\ov g}^{ijk} {\ov \y}_{j \dot  \b} 
{\ov A}_k  
-C_{\a}
{\ov \G}^i  
 {\ov C}_{\dot  \b}
\ee
and using a Fierz transformation plus the definitions 
(\ref{fund1cc}) and (\ref{fund3cc}) this takes the desired form
in (\ref{wetqtgtgerhhrtov}):
\be
\d {\ov W}^i_{\f,\a \dot \b}   
=\pa_{ \g \dot \b }  {\ov \f}^i_{\a} C^{\g} - {\ov \L}^i_{\f, \a} {\ov C}_{\dot \b}  
+ 
2 C_{\a}   
 {\ov g}^{ijk} {\ov A}_j  {\ov \y}_{k \dot \b} 
\ee
Here is the Fierz transformation:
\be
\pa_{\a \dot \b}  \y^{i}_{  \g} {C}^{  \g} 
+ 
  \pa_{\g \dot \b  }   
{ \y}^{i  \g} C_{\a}
=
- {C}_{  \g} \pa_{\a \dot \b}  \y^{i   \g} 
+ 
 C_{\a} \pa_{\g \dot \b  }   
{ \y}^{i  \g} 
\eb
= - \ve_{\a \g} 
{C}^{  \d} \pa_{\d \dot \b}  \y^{i   \g} 
= 
{C}^{  \d} \pa_{\d \dot \b}  \y^{i}_{   \a} 
= 
 \pa_{\d \dot \b}  \ov \f^{i}_{   \a} {C}^{  \d}
\ee
where we use
\be
- \ve_{\a \g} A^{\d} B_{\d}  = A_{\a} B_{\g} - A_{\g} B_{\a} 
\ee
which we verify by
\be
\ve^{\g \a} (- \ve_{\a \g} A^{\d} B_{\d} ) = 
\ve^{\g \a}( A_{\a} B_{\g} - A_{\g} B_{\a} )
= 2 A^{\g} B_{\g}
\ee
See Appendix \ref{fggegghetgrege} for further conventions.
\subsection{Verification of the third transformation}

The third transformation is
\be
\d {\ov \L}^i_{\f,\a }   =
-\d \ov \G^{i}   
C_{ \a} 
\ee
and using  $\d$ defined by  Table 
\ref{physicaltable}, we get
\be
\d {\ov \L}^i_{\f,\a }   =
-\lt (
- \fr{1}{2} \pa_{ \a \dot \b  }       \pa^{ \a \dot \b  }        { A}^{i} 
+  {\ov g}^{ijk}     {\ov  G}_{j k} 
\\
- \pa_{ \d \dot \b } {\ov Y}^{ i \dot \b}    {C}^{\d}   
\rt )
C_{ \a} 
\eb
=
 \fr{1}{2} \pa_{ \g \dot \b  }       \pa^{ \g \dot \b  }        { A}^{i} C_{ \a} 
-  {\ov g}^{ijk}     {\ov  G}_{j k} C_{ \a} 
\\
+ \pa_{ \d \dot \b } {\ov Y}^{ i \dot \b}    {C}^{\d}   
C_{ \a}\eb
=
 \fr{1}{2} \pa_{ \g \dot \b  }       \pa^{ \g \dot \b  }        { A}^{i} C_{ \a} 
-  {\ov g}^{ijk}     {\ov  G}_{j k} C_{ \a} 
\\
+ \pa_{ \d \dot \b } {\ov Y}^{ i \dot \b}    {C}^{\d}   
C_{ \a} 
\la{fweqfwefwefwew}
\ee
and we want to get
\be
\d
 {\ov \L}^i_{\f, \a} = \pa_{ \g \dot \b }  {\ov W}_{\f, \a}^{i\;\; \dot \b}  C^{\g} 
- 
 C_{\a}   
 {\ov g}^{ijk}{\ov  G}_{j k} 
\la{fweqfwefwefwew2}
\ee
where
\be
{\ov W}^i_{\f,\a \dot \b}   =
\lt (
\pa_{\a \dot \b} A^i 
+  C_{\a} \ov Y^i_{\dot  \b}  \rt )
\ee
Substitution in equation (\ref{fweqfwefwefwew2})  yields 
\be
\d
 {\ov \L}^i_{\f, \a} = \pa_{ \g \dot \b }  \lt (
\pa_{\a}^{\;\; \dot \b} A^i 
+  C_{\a} \ov Y^{i  \dot  \b}  \rt ) C^{\g} 
- 
 C_{\a}   
 {\ov g}^{ijk}{\ov  G}_{j k} 
\eb
=    \ve_{\g \a} 
\D A^i C^{\g} 
+  C_{\a} \pa_{ \g \dot \b }   \ov Y^{i  \dot  \b}  C^{\g} 
- 
 C_{\a}   
 {\ov g}^{ijk}{\ov  G}_{j k} 
\eb
=    
\D A^i C_{\a} 
+  C_{\a} \pa_{ \g \dot \b }   \ov Y^{i  \dot  \b}  C^{\g} 
- 
 C_{\a}   
 {\ov g}^{ijk}{\ov  G}_{j k} 
\eb
=
 \fr{1}{2} \pa_{ \g \dot \b  }       \pa^{ \g \dot \b  }        { A}^{i} C_{ \a} 
-  {\ov g}^{ijk}     {\ov  G}_{j k} C_{ \a} 
\\
+ \pa_{ \d \dot \b } {\ov Y}^{ i \dot \b}    {C}^{\d}   
C_{ \a}\ee
So we have shown that equations (\ref{fweqfwefwefwew}) and 
(\ref{fweqfwefwefwew2}) are the same, and we have established that indeed (\ref{fweqfwefwefwew2}) is true. 

Here we have used
\be
\pa_{ \g \dot \b }  
\pa_{\a}^{\;\; \dot \b}  
= \ve_{\g \a}
\D
\ee
which is verified by
\be
\ve^{\g \a}\pa_{ \g \dot \b }  
\pa_{\a}^{\;\; \dot \b}  
= \ve^{\g \a}\ve_{\g \a}
\D = 2 \D =   \pa_{ \g \dot \b  }       \pa^{ \g \dot \b  } 
\ee
consistent with the conventions in Appendix \ref{fggegghetgrege}.

So we have established equation (\ref{fwefwfwefcc}) and its component form (\ref{wetqtgtgerhhrtov}) 
 in detail.   The complex conjugates (\ref{fwefwfwef}) and 
(\ref{wetqtgtgerhhrt}) are therefore also true.

\section{Construction of the full forms for the Simple Pseudosuperfields ${\hat \w}_{(\dot \a_1 \cdots \dot \a_{2m+1})}$ etc.}

\la{qfwqergreg}

 Hence the constraints such as equation 
(\ref{ffweweffwe}) can be seen to result from equation
(\ref{fwefwfwef}) simply by taking the $\q$ and $\ov \q$ independent parts.

So we see that in fact there are chiral and antichiral  pseudosupermultiplets here generated by the forms in subsection
 \ref{fdaffsdfa}, provided that the constraints like 
(\ref{ffweweffwe}) are satisfied. All we need to do is add hats to all the fields in subsection
 \ref{fdaffsdfa} to get the pseudosuperfields rather than the generators. 

We also perform the substitution 
\be
\ov \y \ra {\hat \f}
\ee
to minimize confusion when going from the simple generators to the pseudosuperspace form.

Here, for example, is the composite dotspinor pseudosuperfield 
that corresponds to the generator
(\ref{dotpspinorgeneralfermion}):
\be
 {\hat \w}_{(\dot \a_1 \cdots \dot \a_{2m+1})} = 
\sum_{{\rm Symmetrize}\; \dot \a_r}
\ov f_{(j_1 \cdots j_n)}^{[i_1 \cdots i_{2m+1}]}
{\hat \f}_{i_1 \dot \a_1}  \cdots {\hat \f}_{i_{2m+1} \dot \a_{2m+1}}   {\hat A}^{j_1} \cdots {\hat A}^{j_n}; 
\la{dotpspinorgeneralfermion2super}
\eb
 \; m=0, 1,2 \cdots ,n= 0,1,2 \cdots 
\ee
The `projection' techniques discussed in 
\ci{superspace} are useful in reducing these expressions to components. 

The constraints mean that the resulting simple chiral dotspinor composite pseudosuperfields like ${\hat \w}_{(\dot \a_1 \cdots \dot \a_{2m+1})} $ transform like superfields without the extra terms $-
 g_{ijk} 
{\hat A}^j {\hat A}^k \ov C_{\dot \a}
$  that are in the transformation of the fundamental dotspinor ${\hat \f}_{i\dot \a} $.  That in turn means that their highest component (the coefficient of $\q\cdot \q$) transforms into a total derivitive, which is why they are found in the cohomology space.   The cohomology formed from these pseudosuperfields and generated by the simple generators always has free unsaturated Lorentz spinor indices, and gives rise to certain expressions, the highest component of which is analogous to the $F$ term of a chiral scalar multiplet, in that its variation using  $\d_{\rm BRS}$ is a total derivative.  

Strictly speaking, we have found invariants through the simple generators, but we have not yet shown that they are not boundaries of the operator $\d$. This is necessary to establish that these are really cohomology and not just boundaries.    For given simple examples, it is possible to show this in a simple way, and that is adequate for present purposes.  This is discussed further in section \ref{greghghergherg}.

In fact a careful look at what is going on in \ci{cybersusyI} shows that it does not even matter whether the lepton operators there are in the cohomology space or not.  The result does not depend on that, since the cybersusy action is independently nilpotent and supersymmetry breaking, and really all that the leptonic operators are doing is providing a motivation to look at the cybersusy action.  Of course, it is  important that the leptonic operators do have some physics in them, which relates to the cohomology.

\section{A very important example with a review of the notation and two pedagogical examples verifying that this is a chiral dotted superfield}
\la{asdfasfwerfwefwe2}

Now let us consider a non-trivial, and very important, example of the generator (\ref{dotpspinorgeneralfermion}):
\be
\w_{\dot \a  } =  f_{j}^{i }
\oy^{}_{i \dot \a }   A^{j}
\ee
This yields the superspace version from
(\ref{dotpspinorgeneralfermion2super}):
\be
 {\hat \w}_{\dot \a } = 
  f_{j  }^{i }
{\hat \f}_{i\dot \a}    {\hat A}^{j} 
\ee
and use of the forms for the fundamental dotspinor in equation 
(\ref{weird}) 
and the fundamental scalar from (\ref{qgregerjtjt})
yields
\be
 {\hat \w}_{\dot \a  }(x) = 
\eb
  f_{j  }^{i }
\lt (
\ov \y_{\dot \a i}(y) +
\q^{\b} 
\lt (
\pa_{\b \dot \a} \A_i(y) 
+ \ov C_{\dot \a} Y_{i  \b}(y) \rt )
- \fr{1}{2} \q^{\g} \q_{\g} 
\G_{i}(y)  
\ov C_{\dot \a} \rt )
  \eb
\lt ( A^j(y)
+
\q^{\d} 
\y^j_{\d}(y)
+ \fr{1}{2} \q^{\e} \q_{\e} 
G^j(y) \rt )
\ee
Then projection yields the following components:
\be
\w_{\dot \a  } =  {\hat \w}_{\dot \a  }(x){}_| = f_{j}^{i }
\oy^{}_{i\dot \a }   A^{j}
\ee
\be
W_{\b \dot \a  } = {D}_{\b} {\hat \w}_{\dot \a  }(x){}_| 
\eb
=
  f_{j}^{i }
\lt \{
   \y^{j}_{\b} \oy^{}_{\dot \a i}+ 
\lt (
\pa_{\b \dot \a} \A_i 
+ \ov C_{\dot \a} Y_{i  \b} \rt )A^{j}
\rt \}
\ee
\be
\L_{\dot \a} =
\fr{1}{2} {D}^{\b}{D}_{\b}  {\hat \w}_{\dot \a  }(x){}_|
\eb
= 
 f_j^i
\lt \{
-\G_{i} A^j
\ov C_{\dot \a} 
-
\lt (
\pa_{\b \dot \a} \A_i 
+ \ov C_{\dot \a} Y_{i  \b} \rt )
 \y^{j \b}
+
\oy_{i \dot \a} 
G^{j}\rt \} 
\la{ggrwegegerger}
\ee
where we use
\be
\q^{\b} 
\q^{\d} 
= 
- \fr{1}{2} \ve^{\b \d}
\q^{\e} 
\q_{\e} 
\ee 

\be
\fr{1}{4}
\lt \{ 
{D}^{\b}{D}_{\b}
\q^{\e} 
\q_{\e} \rt \}_|
=1
\ee

Here the symbols are, in the order shown in 
equation (\ref{ggrwegegerger}), defined as follows:
\ben
\item
The technique of projection used here is explained in 
\ci{superspace}.  It is the fastest way to go from a product of superfields to the corresponding component expression.  The notation ${D}_{\b} {\hat \w}_{\dot \a  }(x){}_| $ means  `evaluate the expression and then set $\q = \ov \q = 0$'.
\item
$\L_{\dot \a} 
$ is our standard notation for the highest weight component of a chiral dotspinor multiplet--note that the spinor index on $\L_{\dot \a} $ is a dotted spinor index.
\item
The tensor $f_j^i$ is a numerical tensor contracted with flavour or internal indices $i$
\item
$\G_{i}$ is the Zinn-Justin source for the variation of the scalar field $\d A^i$,
\item
 $A^j$ is the scalar field in the chiral multiplet,
\item
$\ov C_{\dot \a}$ is the complex conjugate  commuting Weyl spinor spacetime independent supersymmetry ghost. 
($C_{\a}$ is the commuting Weyl spinor spacetime independent supersymmetry ghost)

\item
 $\A_j$ is the complex conjugate scalar field in the complex conjugate antichiral multiplet,
\item
$Y_{i  \b}$ is the Zinn-Justin source for the variation of the spinor field $\d \y^{i \b}$,
\item
 $\y^j_{\a}$ is the spinor field in the chiral multiplet,
\item
 $\oy_{j \dot \a}$ is the  complex conjugate spinor field in the complex conjugate antichiral multiplet,
\item
$G^{j}$ is a composite term which arises from integration of the $F$ auxiliary field, and has the form
\[
G^{j}= \lt ( 
{\ov g}^{jlq} \A_l \A_q
+
{\ov Y}^j_{\dot  \b} {\ov C}^{\dot \b } 
\rt )
\]
where $ {\ov g}^{jlq}$ is the complex conjugate of the tensor in the superpotential.
\item
${\ov Y}^i_{\dot  \b}$ is the complex conjugate Zinn-Justin source for the variation of the spinor field $\d \oy_{i}^{ \dot \b}$,
\item
The tensor $f_j^i$ is required to satisfy the symmetrization constraint:
\be
  f^s_{(i}  g_{jk)s} \equiv
  f^s_{i}  g_{jks}  +
  f^s_{j}  g_{kis}  +
  f^s_{k}  g_{ij s}  
=0 
\la{fgfgwefgwefg2}
\ee
where $ g_{ijk}$ is the tensor in the superpotential.

\een

So we see that the generators 
(\ref{dotpspinorgeneralfermion})
and 
(\ref{dotpspinorgeneralboson})
 are actually the lowest components of the dotted chiral spinor superfields that they generate. Similarly  
(\ref{undotpspinorgeneralfermion})
and
(\ref{undotpspinorgeneralboson}) are actually the lowest components of the undotted chiral spinor superfields that they generate.  The higher components involve the Zinn sources. 

\section{Verification of the  Transformations for the full dotspinor ${\hat \w}_{\dot \a } = 
  f_{j  }^{i }
{\hat \f}_{i\dot \a}    {\hat A}^{j} 
$}

So we have
\be
\w_{\dot \a  } =  {\hat \w}_{\dot \a  }(x){}_| = f_{j}^{i }
\oy^{}_{i\dot \a }   A^{j}
\ee
\be
W_{\b \dot \a  } = {D}_{\b} {\hat \w}_{\dot \a  }(x){}_| 
=
  f_{j}^{i }
\lt \{
   \y^{j}_{\b} \oy^{}_{\dot \a i}+ 
\lt (
\pa_{\b \dot \a} \A_i 
+ \ov C_{\dot \a} Y_{i  \b} \rt )A^{j}
\rt \}
\la{gghghegh}
\ee
\be
\L_{\dot \a} =
\fr{1}{2} {D}^{\b}{D}_{\b}  {\hat \w}_{\dot \a  }(x){}_|
\eb
= 
    f_j^i
\lt \{
-\G_{i} A^j
\ov C_{\dot \a} 
+
\lt (
\pa_{\b \dot \a} \A_i 
+ \ov C_{\dot \a} Y_{i  \b} \rt )
 \y^{j \b}
+
\oy_{i \dot \a} 
G^{j}\rt \} 
\ee

For clarity, we include two exhaustive verifications that this combination does indeed transform as a chiral dotted spinor superfield under the transformations induced by 
Table \ref{physicaltable}:

\subsection{Verification of the First Transformation for the full dotspinor ${\hat \w}_{\dot \a } = 
  f_{j  }^{i }
{\hat \f}_{i\dot \a}    {\hat A}^{j} 
$}

Let us look at the variation of the first term above: 
\be
\d \w_{\dot \a  } =   f_{j}^{i}
\lt ( \d \oy^{}_{\dot \a i}   A^{j}
- \oy^{}_{\dot \a i} \d  A^{j} \rt )
\ee
\be
=  f_{j}^{i }
\lt ( 
\lt [\pa_{ \a \dot \a }  {\ov A}_{i} {C}^{\a}  
+ 
{\ov C}_{\dot \a}   
{\ov G}_{i}
 \rt ] A^{j}
- \oy^{}_{\dot \a i} \y^{j}_{  \b} {C}^{  \b}  \rt )
\ee
where we use Table \ref{physicaltable}, and then use
\be
{\ov G}_{i} 
=-\lt ( 
{g}_{ilq} A^l A^q
+
{Y}_i^{  \b} {C}_{\b } 
\rt )
\ee
from Table \ref{compterms}. 
Putting these together yields
\be
\d \w_{\dot \a  } =  f_{j}^{i}
\lt ( \d \oy^{}_{\dot \a i}   A^{j}
- \oy^{}_{\dot \a i} \d  A^{j} \rt )
\ee
\be
= f_{j}^{i }
\lt ( 
\lt [\pa_{ \a \dot \a }  {\ov A}_{i} {C}^{\a}  
- 
{\ov C}_{\dot \a}   
\lt ( 
{g}_{ilq} A^l A^q
+
{Y}_i^{  \b} {C}_{\b } 
\rt )
 \rt ] A^{j}
- \oy^{}_{\dot \a i} \y^{j}_{  \b} {C}^{  \b}  \rt )
\ee
\be
= f_{j}^{i }
\lt \{
 \lt (\pa_{ \a \dot \a }  {\ov A}_{i} 
+ 
{\ov C}_{\dot \a}   
{Y}_{i \a} 
 \rt ) A^{j}
+\y^{j}_{  \a}  \oy^{}_{\dot \a i}  
\rt \}
{C}^{\a}  
\ee
where we have used the constraint:
\be
f_{j}^{i }
{g}_{ilq} A^l A^q
A^{j}
=0
\ee
Now  since
\be
W_{\b \dot \a_1  } =  f_{j}^{i }
\lt \{
-\oy^{}_{\dot \a_1 i}   \y^{j}_{\b}+ 
\lt (
\pa_{\b \dot \a} \A_i 
+ \ov C_{\dot \a} Y_{i  \b} \rt )A^{j}
\rt \}\ee
we see that this is
\be
\d \w_{\dot \a  }=
W_{\a \dot \a  }{C}^{\a}  
\ee
So this term is transforming as a chiral dotted spinor superfield.

We shall skip the second term, and proceed to the third one:

\subsection{Verification of the Third Transformation for the full dotspinor ${\hat \w}_{\dot \a } = 
  f_{j  }^{i }
{\hat \f}_{i\dot \a}    {\hat A}^{j} 
$}
So this is:
\be
\d \L_{\dot \a} =
\eb
    f_j^i
 \lt \{
-\d\G_{i} A^j
\ov C_{\dot \a} 
-
\lt (
\pa_{\b \dot \a} \d \A_i 
+ \ov C_{\dot \a} \d Y_{i  \b} \rt )
\y^{j \b}
+
\d \oy_{i \dot \a} 
G^{j}\rt \} 
\eb
 +  f_j^i
 \lt \{
+\G_{i} \d A^j
\ov C_{\dot \a} 
-
\lt (
\pa_{\b \dot \a} \A_i 
+ \ov C_{\dot \a} Y_{i  \b} \rt )
\d \y^{j \b}
-
\oy_{i \dot \a} 
\d G^{j}\rt \} 
\ee
Now using    Table \ref{physicaltable}, this becomes:
\be
\d \L_{\dot \a} =
\eb
    f_j^i
 \lt \{
-\lt (
 - \fr{1}{2} \pa_{ \a \dot \b  }       \pa^{ \a \dot \b  }        {\ov  A}_{i} 
 + g_{ijk} { G}^{jk}
-  \pa_{ \a \dot \b } Y_{i}^{ \a}    {\ov C}^{\dot \b}   
\rt )
 A^j
\ov C_{\dot \a} 
\ebp
-
\lt [
\pa_{\b \dot \a} {\ov \y}_{i  \dot \b} {\ov C}^{ \dot  \b} 
+ \ov C_{\dot \a} \lt ( +
  \pa_{\b \dot \b  }   
{\ov \y}_{i}^{  \dot \b}
 +
2 g_{ijk}  \y^{j}_{ \b} A^k    
-
\G_i  
 {C}_{  \b}
 \rt )
 \rt ]
 \y^{j \b}
\ebp
\lt.
+
\lt ( \pa_{ \a \dot \a }  {\ov A}_{i} {C}^{\a}  
+ 
{\ov C}_{\dot \a}   
{\ov G}_{i}
 \rt )
G^{j}\rt \} 
\ebp
\rt. +  f_j^i
 \lt \{
+\G_{i}  \y^{j}_{  \b} {C}^{  \b} 
\ov C_{\dot \a} 
\ebp
-
\lt (
\pa_{\b \dot \a} \A_i 
+ \ov C_{\dot \a} Y_{i  \b} \rt )
\lt ( -\pa^{ \b \dot \b }  A^{j} {\ov C}_{\dot \b}  
+ 
C^{\b}   
G^j
 \rt )
\ebp
-
\oy_{i \dot \a} 
  \pa_{\a \dot \b}   \y^{j \a} {\ov C}^{\dot \b} 
\rt \} 
\ee
which is
\be
\d \L_{\dot \a} =
\eb
    f_j^i
 \lt \{
+ \fr{1}{2} \pa_{ \a \dot \b  }       \pa^{ \a \dot \b  }        {\ov  A}_{i}  A^j
\ov C_{\dot \a} 
 - g_{ijk} { G}^{jk} A^j
\ov C_{\dot \a} 
+  \pa_{ \a \dot \b } Y_{i}^{ \a}    {\ov C}^{\dot \b}   
 A^j
\ov C_{\dot \a} 
\ebp
-
\pa_{\b \dot \a} {\ov \y}_{i  \dot \b} {\ov C}^{ \dot  \b} 
 \y^{j \b}
\ebp
-    \pa_{\b \dot \b  }   
{\ov \y}_{i}^{  \dot \b}
\ov C_{\dot \a} \y^{j \b}
 -
2 g_{ijk}  \y^{j}_{ \b} A^k  \ov C_{\dot \a} \y^{j \b}  
+
\G_i  
 {C}_{  \b}
\ov C_{\dot \a} \y^{j \b}
\ebp
\lt.
+
  \pa_{ \a \dot \a }  {\ov A}_{i} {C}^{\a}  
G^{j}+ 
{\ov C}_{\dot \a}   
{\ov G}_{i}
G^{j}\rt.
+\G_{i}  \y^{j}_{  \b} {C}^{  \b} 
\ov C_{\dot \a} 
\ebp
+
\pa_{\b \dot \a} \A_i \pa^{ \b \dot \b }  A^{j} {\ov C}_{\dot \b}  
+ \ov C_{\dot \a} Y_{i  \b}
\pa^{ \b \dot \b }  A^{j} {\ov C}_{\dot \b}  
\ebp
- 
\pa_{\b \dot \a} \A_i 
C^{\b}   
G^j
- \ov C_{\dot \a} Y_{i  \b} 
C^{\b}   
G^j
\ebp
-
\oy_{i \dot \a} 
  \pa_{\a \dot \b}   \y^{j \a} {\ov C}^{\dot \b} 
\rt \} 
\ee
Now collect like terms
\be
\d \L_{\dot \a} =
\eb
    f_j^i
 \lt \{
+ \fr{1}{2} \pa_{ \a \dot \b  }       \pa^{ \a \dot \b  }        {\ov  A}_{i}  A^j
\ov C_{\dot \a} 
+
\pa_{\b \dot \a} \A_i \pa^{ \b \dot \b }  A^{j} {\ov C}_{\dot \b}  
\ebp
 - g_{ijk} { G}^{jk} A^j
\ov C_{\dot \a} 
 -
2 g_{ikl}  \y^{k}_{ \b} A^l  \ov C_{\dot \a} \y^{j \b}  
\ebp
+ 
{\ov C}_{\dot \a}   
{\ov G}_{i}
G^{j}
- \ov C_{\dot \a} Y_{i  \b} 
C^{\b}   
G^j
\ebp
+  \pa_{ \a \dot \b } Y_{i}^{ \a}    {\ov C}^{\dot \b}   
 A^j
\ov C_{\dot \a} 
+ \ov C_{\dot \a} Y_{i  \b}
\pa^{ \b \dot \b }  A^{j} {\ov C}_{\dot \b}  
\ebp
-
\pa_{\b \dot \a} {\ov \y}_{i  \dot \b} {\ov C}^{ \dot  \b} 
 \y^{j \b}
-    \pa_{\b \dot \b  }   
{\ov \y}_{i}^{  \dot \b}
\ov C_{\dot \a} \y^{j \b}
-
\oy_{i \dot \a} 
  \pa_{\a \dot \b}   \y^{j \a} {\ov C}^{\dot \b} 
\ebp
+
  \pa_{ \a \dot \a }  {\ov A}_{i} {C}^{\a}  
G^{j}
- 
\pa_{\b \dot \a} \A_i 
C^{\b}   
G^j
\ebp
+
\G_i  
 {C}_{  \b}
\ov C_{\dot \a} \y^{j \b}
+\G_{i}  \y^{j}_{  \b} {C}^{  \b} 
\ov C_{\dot \a} 
\rt \} 
\ee
Now the last four terms cancel in pairs and we have
\be
\d \L_{\dot \a} =
\eb
    f_j^i
 \lt \{
+ \fr{1}{2} \pa_{ \a \dot \b  }       \pa^{ \a \dot \b  }        {\ov  A}_{i}  A^j
\ov C_{\dot \a} 
+
\pa_{\b \dot \a} \A_i \pa^{ \b \dot \b }  A^{j} {\ov C}_{\dot \b}  
\ebp
 - g_{ikl} { G}^{kl} A^j
\ov C_{\dot \a} 
 -
2 g_{ikl}  \y^{k}_{ \b} A^l  \ov C_{\dot \a} \y^{j \b}  
\ebp
+ 
{\ov C}_{\dot \a}   
{\ov G}_{i}
G^{j}
- \ov C_{\dot \a} Y_{i  \b} 
C^{\b}   
G^j
\ebp
+  \pa_{ \a \dot \b } Y_{i}^{ \a}    {\ov C}^{\dot \b}   
 A^j
\ov C_{\dot \a} 
+ \ov C_{\dot \a} Y_{i  \b}
\pa^{ \b \dot \b }  A^{j} {\ov C}_{\dot \b}  
\ebp
-
\pa_{\b \dot \a} {\ov \y}_{i  \dot \b} {\ov C}^{ \dot  \b} 
 \y^{j \b}
-    \pa_{\b \dot \b  }   
{\ov \y}_{i}^{  \dot \b}
\ov C_{\dot \a} \y^{j \b}
-
\oy_{i \dot \a} 
  \pa_{\a \dot \b}   \y^{j \a} {\ov C}^{\dot \b} 
\rt \} 
\ee
Now we can write this as
\be
\d \L_{\dot \a} =
\eb
    f_j^i
 \lt \{
+ \fr{1}{2} \pa_{ \a \dot \b  }       \pa^{ \a \dot \b  }        {\ov  A}_{i}  A^j
\ov C_{\dot \a} 
+
\pa^{ \b \dot \b }
\lt (
\pa_{\b \dot \a}  \A_i   A^{j} {\ov C}_{\dot \b}  
\rt )
-
\pa^{ \b \dot \b }
\pa_{\b \dot \a}  \A_i   A^{j} {\ov C}_{\dot \b}  
\ebp
 - g_{ikl} \lt ( 2 A^k G^l -\y^{k \g} \y^{l}_{ \g} \rt )
 A^j
\ov C_{\dot \a} 
 -
2 g_{ikl}  \y^{k}_{ \b} A^l  \ov C_{\dot \a} \y^{j \b}  
\ebp
+ 
{\ov C}_{\dot \a}   
\lt ( - Y_{i}^{  \b} 
C_{\b}   - g_{ikl} A^k A^l
\rt )
G^{j}
- \ov C_{\dot \a} Y_{i  \b} 
C^{\b}   
G^j
\ebp
+  \pa_{ \a \dot \b } \lt (
Y_{i}^{ \a}    {\ov C}^{\dot \b}   
 A^j
\ov C_{\dot \a} 
\rt )
\ebp
-
\pa_{\b \dot \a} {\ov \y}_{i  \dot \b} {\ov C}^{ \dot  \b} 
 \y^{j \b}
-    \pa_{\b \dot \b  }   
{\ov \y}_{i}^{  \dot \b}
\ov C_{\dot \a} \y^{j \b}
-
\oy_{i \dot \a} 
  \pa_{\a \dot \b}   \y^{j \a} {\ov C}^{\dot \b} 
\rt \} 
\ee
and this simplifies to 
\[
\d \L_{\dot \a} =
    f_j^i
 \lt \{
\pa^{ \b \dot \b }
\lt (
\pa_{\b \dot \a}  \A_i   A^{j} {\ov C}_{\dot \b}  
\rt )
\rt.
\]
\[
 - g_{ikl}  2 A^k G^l   
 A^j
\ov C_{\dot \a} 
- 
{\ov C}_{\dot \a}   
 g_{ikl} A^k A^l
G^{j}
\]\[
 + g_{ikl} \y^{k \g} \y^{l}_{ \g} 
 A^j
\ov C_{\dot \a} 
 -
2 g_{ikl}  \y^{k}_{ \b} A^l  \ov C_{\dot \a} \y^{j \b}  
\]\[
+  \pa_{ \a \dot \b } \lt (
Y_{i}^{ \a}    {\ov C}^{\dot \b}   
 A^j
\ov C_{\dot \a} 
\rt )
\]\be
\lt.
-
\pa_{\b \dot \a} {\ov \y}_{i  \dot \b} {\ov C}^{ \dot  \b} 
 \y^{j \b}
-    \pa_{\b \dot \b  }   
{\ov \y}_{i}^{  \dot \b}
\ov C_{\dot \a} \y^{j \b}
-
\oy_{i \dot \a} 
  \pa_{\a \dot \b}   \y^{j \a} {\ov C}^{\dot \b} 
\rt \} 
\la{qgreghhrhwhrt}
\ee
From (\ref{fqwertgreger}), we expect this to be 
\be
\d
 { \L}_{ \dot \a} =  \pa_{ \b \dot \g }  { W}_{ \;\;\dot \a}^{ \b}  \ov C^{\dot \g} 
 + \x^{\g \dot \d} \partial_{\g \dot \d}  { \L}_{ \dot \a}
\ee
and we can ignore the $\x^{\g \dot \d}$ term here. 
From (\ref{gghghegh}), we have
\be
W_{\b \dot \a  }  
=
  f_{j}^{i }
\lt \{
   \y^{j}_{\b} \oy^{}_{\dot \a i}+ 
\lt (
\pa_{\b \dot \a} \A_i 
+ \ov C_{\dot \a} Y_{i  \b} \rt )A^{j}
\rt \}
\ee
So we should find that
\be
\d
 { \L}_{ \dot \a} =  \pa_{ \b \dot \g }    f_{j}^{i }
\lt \{
   \y^{j \b} \oy^{}_{\dot \a i}+ 
\lt (
\pa^{\b}_{\;\; \dot \a} \A_i 
+ \ov C_{\dot \a} Y_{i}^{  \b} \rt )A^{j}
\rt \}
 \ov C^{\dot \g} 
\ee
is the same as (\ref{qgreghhrhwhrt}). 
For this to be true we require the following
\ben
\item
Firstly, the following should follow from the constraint:
\be
0 =
    f_j^i
 \lt \{
 - g_{ikl}  2 A^k G^l   
 A^j
\ov C_{\dot \a} 
- 
{\ov C}_{\dot \a}   
 g_{ikl} A^k A^l
G^{j}
\ebp
 + g_{ikl} \y^{k \g} \y^{l}_{ \g} 
 A^j
\ov C_{\dot \a} 
 -
2 g_{ikl}  \y^{k}_{ \b} A^l  \ov C_{\dot \a} \y^{j \b}  
\rt \} 
\la{gregergerg}
\ee
\item
Secondly, the following should follow from a Fierz transformation:
\be
  \pa_{ \b \dot \g }    f_{j}^{i }
\lt \{
   \y^{j \b} \oy^{}_{\dot \a i} 
\rt \}
 \ov C^{\dot \g} 
\la{qreggherhrthrthrth1}
\eb
=    f_j^i
 \lt \{
-
\pa_{\b \dot \a} {\ov \y}_{i  \dot \b} {\ov C}^{ \dot  \b} 
 \y^{j \b}
-    \pa_{\b \dot \b  }   
{\ov \y}_{i}^{  \dot \b}
\ov C_{\dot \a} \y^{j \b}
-
\oy_{i \dot \a} 
  \pa_{\a \dot \b}   \y^{j \a} {\ov C}^{\dot \b} 
\rt \} 
\la{rewrwerwerwe}
\ee
\item
Thirdly, we should have
\be
  \pa_{ \b \dot \g }    f_{j}^{i }
\lt \{
\pa^{\b}_{\;\; \dot \a} \A_i A^{j}
+ \ov C_{\dot \a} Y_{i}^{  \b}  A^{j}
\rt \}
 \ov C^{\dot \g} 
\eb=
    f_j^i
\pa^{ \b \dot \b }
\lt (
\pa_{\b \dot \a}  \A_i   A^{j} {\ov C}_{\dot \b}  
\rt )
+ f_j^i \pa_{ \a \dot \b } \lt (
Y_{i}^{ \a}    {\ov C}^{\dot \b}   
 A^j
\ov C_{\dot \a} \rt )
\ee
\een
These are all simple exercizes.
 For example, from  (\ref{gregergerg}), we have:
\be
    f_j^i
 g_{ikl} \y^{k \g} \y^{l}_{ \g} 
 A^j
\ov C_{\dot \a} 
 -
    f_j^i
2 g_{ikl}  \y^{k}_{ \b} A^l  \ov C_{\dot \a} \y^{j \b}  
\eb
=
 -   f_j^i
 g_{ikl} \y^{k}_{ \g} \y^{l  \g} 
 A^j
\ov C_{\dot \a} 
 -
    f_j^i
 g_{ikl}  \y^{k}_{ \b} A^l  \ov C_{\dot \a} \y^{j \b}  
 -
    f_j^i
 g_{ikl}  \y^{j}_{ \b} A^l  \ov C_{\dot \a} \y^{k \b}  
\eb
=
 -  
\lt \{
 f_j^i
 g_{ikl}  
 +
    f_k^i
 g_{ijl}   
  +  f_l^i
 g_{ijk} \rt \}
 \y^{j}_{ \b}  \y^{k \b}  A^l  \ov C_{\dot \a}  
=0
\ee
Also, we have from (\ref{rewrwerwerwe})
\be
    f_j^i
 \lt \{
-
\pa_{\b \dot \a} {\ov \y}_{i  \dot \b} {\ov C}^{ \dot  \b} 
 \y^{j \b}
-    \pa_{\b \dot \b  }   
{\ov \y}_{i}^{  \dot \b}
\ov C_{\dot \a} \y^{j \b}
-
\oy_{i \dot \a} 
  \pa_{\a \dot \b}   \y^{j \a} {\ov C}^{\dot \b} 
\rt \} 
\eb
=    f_j^i
 \lt \{
+
\pa_{\b \dot \a} {\ov \y}_{i}^{  \dot \b} {\ov C}_{ \dot  \b} 
 \y^{j \b}
-    \pa_{\b \dot \b  }   
{\ov \y}_{i}^{  \dot \b}
\ov C_{\dot \a} \y^{j \b}
-
\oy_{i \dot \a} 
  \pa_{\a \dot \b}   \y^{j \a} {\ov C}^{\dot \b} 
\rt \} 
\eb
=    f_j^i
 \lt \{
-
\ve_{\dot \a \dot \b}
\pa_{\b}^{\;\; \dot \d} {\ov \y}_{i}^{  \dot \b} {\ov C}_{ \dot  \d} 
 \y^{j \b}
-
\oy_{i \dot \a} 
  \pa_{\a \dot \b}   \y^{j \a} {\ov C}^{\dot \b} 
\rt \} 
\eb
=    f_j^i
 \lt \{
+
\pa_{\b}^{\;\; \dot \d} {\ov \y}_{i  \dot \a} {\ov C}_{ \dot  \d} 
 \y^{j \b}
-
\oy_{i \dot \a} 
  \pa_{\a \dot \b}   \y^{j \a} {\ov C}^{\dot \b} 
\rt \} 
\eb
=    f_j^i
 \lt \{
-
\pa_{\a  \dot \b} {\ov \y}_{i  \dot \a} {\ov C}^{ \dot  \b} 
 \y^{j \a}
-
\oy_{i \dot \a} 
  \pa_{\a \dot \b}   \y^{j \a} {\ov C}^{\dot \b} 
\rt \} 
\eb
= -   f_j^i
\pa_{\a  \dot \b} 
 \lt \{
{\ov \y}_{i  \dot \a} {\ov C}^{ \dot  \b} 
 \y^{j \a}
\rt \} 
= +   f_j^i
\pa_{\a  \dot \b} 
 \lt \{
 \y^{j \a}
{\ov \y}_{i  \dot \a}  \rt \} {\ov C}^{ \dot  \b}
\ee
which is the expression (\ref{qreggherhrthrthrth1}), as required.
So we have established 
that the third transformation is  correct.

\section{Cohomology}

\subsection{Boundaries}

\la{greghghergherg}

In order to establish that the dotspinors form cohomology objects we need to know that they are not boundaries.

Actually it is the integral of the highest component of any dotspinor that is in the cohomology space.  For example
\be
\int d^4 x\; \L_{\dot \a}
\eb
 =
\int d^4 x\;
    f_j^i
\lt \{
-\G_{i} A^j
\ov C_{\dot \a} 
+
\lt (
\pa_{\b \dot \a} \A_i 
+ \ov C_{\dot \a} Y_{i  \b} \rt )
 \y^{j \b}
+
\oy_{i \dot \a} 
G^{j}\rt \} 
\eb
\in {\cal H}
\ee

We know that this transforms to a total derivative, and so
\be
\d \int d^4 x\; \L_{\dot \a} 
=\int d^4 x\; 
\pa_{ \b \dot \g }  { W}_{ \;\;\dot \a}^{ \b}  \ov C^{\dot \g} =
0
\ee
But to establish that this is really in the cohomology space we need to also show that it is not a boundary.  We need to show that there is no local polynomial $ B_{\dot \a}$ such that:
\be
\int d^4 x\; \L_{\dot \a} = \d \int d^4 x\; B_{\dot \a}
\ee
To do this is not hard.  There are only the following possibilities:
\be
\int d^4 x\;   B_{\dot \a}=
\int d^4 x\;  \lt \{ e_{1,ij} \ov Y^i_{\dot \a} A^j
+e_{2,i}^{j} \ov Y^i_{\dot \a} \A_j  \rt \}
\ee
and these do not work.
The possibilities are very limited because
\ben
\item
The dimension of the integrand must be $3 \fr{1}{2}$;
\item
The ghost number of the integrand must be minus one;
\item
The integrand must have one unsaturated dotted spinor index;
\item
The integrand must not be a total derivative; and 
\item
The integrand must be local.
\een

To prove this in general for the simple dotspinors is probably not very difficult, but it is not essential for now, since we are only using low dimensional examples for the leptonic dotspinors in this series of four papers, and the boundaries are so different from the dotspinors that it is obvious that the dotspinors are in the cohomology space.  The baryons will require a little more work along these lines.

\subsection{The effects of the rest of the cohomology}

The simple generators do not generate all the operators in the cohomology space of the operator in   Table \ref{physicaltable}.
To find all the cohomology requires one to use spectral sequences along the lines of \ci{bigpaper}.  There are still many unsolved problems there.   However it is evident from what is known that there are infinite series of operators with ghost charge zero, one, two... that generalize the simple generators and provide operators in the cohomology space.  What can be expected to arise is a generalization of  the results in 
\ci{dixminram}.  That paper shows that when one includes derivatives  in the operators $\y, \oy, A,\A$ that are used to generate the simple generators, one obtains more cohomology subject to more symmetrization conditions.  Those operators will also be subject to constraints like the ones for the simple generators, and there will be generalized pseudosuperfields to generate the full expressions.

In addition it should be noted that if one finds an operator of physical interest, say 
$ {\hat \w}_{\dot \a } = 
  f_{j  }^{i }
{\hat \f}_{i\dot \a}    {\hat A}^{j} 
$, then there are an infinite number of other operators 
even within the realm of the simple dotspinors, that also would have the same physical interest, and that would also solve the constraints, namely
$ {\hat \w}_{\dot \a } = 
  f_{j  }^{i }
{\hat \f}_{i\dot \a}    {\hat A}^{j} F[{\hat A}]
$
where $F[{\hat A}]$ is any polynomial in ${\hat A}$ that has the quantum numbers of the Lagrangian (except for its mass quantum number).  

But on the other hand, such multiplication by a scalar can be absorbed into the transformation that was contemplated in 
\ci{cybersusyI} where the effective fields were defined, so it would probably not change the results of \ci{cybersusyI}  very much by including this generalization.

The generalization of the simple fields to include derivatives could be expected to make a change however.  But it is a change that could be expected to be of order 
$\lt ( \fr{\rm Momentum}{\rm Mass}\rt )^n$ and therefore suppressed in effect.

Of course, one also needs to generalize Table \ref{physicaltable} to include supersymmetric gauge theory.  This needs a separate treatment of course.  In the next paper \ci{cybersusyIII}, this will be discussed a little more.  The main effect will be to single out gauge invariant specimens of the simple dotspinors, except that we can allow them to have non-zero U(1) charge given our interest in the physics. 

So the conclusion is that the results of the first paper of this series \ci{cybersusyI}, do deserve to be taken as a serious first approximation to  supersymmetry breaking, even when one takes all the cohomology into account.

\section{Supersymmetric Standard Model }
In the next  paper  in this series \ci{cybersusyIII},   it will be shown that these constraint equations have solutions in the massless Supersymmetric Standard Model (SSM), and that these composite chiral dotted spinors (and their complex conjugates) describe interesting composite particles in the massless  SSM, including supersymmetric versions of the familiar hadrons.

It is worth noting the following:
\ben
\item
It appears that the composite chiral dotspinors do not arise when one uses superspace formalism, because they depend essentially on the presence of the Zinn sources, and on the integration of the auxiliary fields. 
This arises because superspace is so implicit and because the chiral superfields are constrained.  Furthermore, non-linear terms, with the  auxiliary fields not integrated, also appear to require manipulation that is equivalent to integration of the auxiliary fields. 
\item
However superspace reappears as shown above, in a constrained way. 
But note that in superspace for the Wess Zumino model, one is not led to invent anything like ${\hat \f}_{i \dot \a} $ in  (\ref{weird}).
\item
There is nothing comparable to this in non-supersymmetric theories, such as the standard model without supersymmetry.
In those theories the Zinn Sources do not play such an important role in forming new composite invariants as they do in (\ref{ggrwegegerger}), and there is nothing comparable to the constraint (\ref{fgfgwefgwefg2}), except   invariance under the gauge group, which is really quite different. 

\item In the first paper of this series \ci{cybersusyI}, we used the fact that  these composite chiral dotspinors lend themselves to the formation of an effective action for various parts of the standard supersymmetric model.

\item
In the next paper of this series \ci{cybersusyIII}, we will see that when one breaks the gauge symmetry spontaneously in the usual way, the effective actions describe a model for broken supersymmetry which arises from the mixing of the usual observable supermultiplets (like the electron) with new composite supermultiplets described by these composite dotspinors.  
\een

\section{The operator $d_3$ and the constraint equations}

\la{dthreechap}

\subsection{The Superpotential Operator $d_3$}

Now that we have the form of $\d_{BRS}$ explicitly, we can have a more complete discussion of the material in section \ref{fdaffsdfa}.

In this paper we will continue to consider only the simple situation where there are no  derivatives, and we will only discuss the action of the following operator:
\be
d_{3 }  
=  
C_{\a} {\ov g}^{ijk} \A_j \A_k 
\y_{\a}^{i \dag} + \oC_{\dot \a}{  g}_{ijk} A^j A^k 
\oy_{ \dot \a}^{i  \dag}    \ee
on objects in the simple subspace described in section 
\ref{fdaffsdfa}.

So, for example, consider the situation where we have an expression of the form:
\be
\A_{(\a \b)}= f_{[ij]}^{pqr}
 \y^i_{(\a} \y^j_{\b)} \A_p \A_q \A_r
\la{qfewfwefw}
\ee

Now we get
\be
d_{3}  
 \A_{(\a \b)}
\eb
=  
 C_{\a}
   {\ov g}^{ijk} \A_j \A_k 
 \y_{\a}^{i \dag} 
f_{[ij]}^{pqr}
\y^i_{(\a} \y^j_{\b)} \A_p \A_q \A_r
\eb
= 2 C_{(\a} \y^j_{\b)}
f_{[ij]}^{pqr}{\ov g}^{ist} \A_s \A_t 
  \A_p \A_q \A_r
\ee
and so the constraint equation is
\be
f_{[ij]}^{(pqr}{\ov g}^{st)i}=0
\la{qgfgwfgw}
\la{qwffwefwefwfe}
\ee
It may seem strange that the SSM has solutions to these equations, given that equation (\ref{qwffwefwefwfe}) is all about symmetrization.  The reason that the SSM affords solutions is that the SSM has a direct product structure of group indices with colour, isospin and hypercharge and it also has three flavours, and it has different representations for the left and right chiralities. In other words, some of the fields in the SSM  
have multiple indices, and the different fields have different numbers of indices, as is well known.  This means that symmetrization can be achieved by double antisymmetrization in the multiple indices, etc.  It will be seen that this is how the SSM comes up with solutions to these symmetrization constraints. 

This also means that grand unified theories  confront some new issues when one looks for solutions of the constraints for such theories. The solutions to the constraints are tightly bound up with the field content.  

\subsection{Invariance of the Superpotential}

There is another way to look at the constraint 
equations.  Consider the operator above in equation 
(\ref{qfewfwefw}):
\be
\A_{(\a \b)}= f_{[ij]}^{pqr}
 \y^i_{(\a} \y^j_{\b)} \A_p \A_q \A_r
\ee
and let us construct the new related operator:
\be
\ov L_{f (\a \b)}  =
f_{[ij]}^{pqr} 
C_{(\a} \y^j_{\b)}    \A_p A_q \A_r
\fr{\pa}{\pa \A_i}
\ee
or more simply the operator
\be
\ov L_{j}  =
f_{[ij]}^{pqr} 
  \A_p A_q \A_r
\fr{\pa}{\pa \A_i}
\ee
operating on the superpotential:
\be
{\ov {\cal Y}}=  {\ov g}^{pqr}\A_p \A_q \A_r 
\ee
Observe that  the constraint (\ref{qgfgwfgw}) can also be obtained as follows:
\be
\ov L_{j}  {\ov {\cal Y}} =0
\Ra f_{[ij]}^{pqr}{\ov g}^{ist} \A_s \A_t 
  \A_p A_q \A_r
=0
\Ra
f_{[ij]}^{(pqr}{\ov g}^{st)i}=0
\ee
From this point of view the constraint can be viewed as an invariance of the superpotential, with invariance operator 
$\ov L_{j} $. 
The set of all eligible operators $\ov L_{j}$ forms a Lie algebra of invariances of the superpotential. 
Moreover, for any $\ov L_{j}$  which generates an invariance 
of the superpotential, we can construct a dotspinor in the cohomology space.

 Similarly for the  complex conjugate, we have:
\be
\ov L^{j}  =
\ov f^{[ij]}_{pqr} 
 A^p A^q A^r
\fr{\pa}{\pa A^i}
\ee
operating on
\be
{\cal Y} 
=  {g}_{ijk} A^j A^k 
  A^i
\ee
The set of all eligible operators $L^{j}$ forms the complex conjugate Lie algebra of invariances of the complex conjugate superpotential.

This observation may yield some
insight  into the constraint equations for an arbitrary superpotential, but it needs further work.
This implies something about the SSM too, but it is not clear to the author what that is.

\section{Conclusion}

In this paper we have derived the BRS transformations for the massless chiral Wess Zumino model and explained how a large set of simple dotspinors and their complex conjugates fit into the cohomology space.    The constraint equations that come from the operator $d_3$ have been explained and illustrated. 
The various dotspinors in their expanded form in terms of components have been exhibited, and  some examples of the transformations have been illustrated in excruciating detail.  
Construction of component forms for the dotspinors has been illustrated using projection on pseudosuperfields. 
It has been shown  that the resulting form of the polynomials in the cohomology space arise from the highest dimension $\q \cdot \q$ components of the dotspinors. For low dimensional examples, the possible boundaries have been examined, and it has been shown  and that  the dotspinor cohomology are not boundaries of the form $\d B$ for any local poynomial $B$. The relationship between simple dotspinor cohomology and the Lie algebra of invariances of the superpotential has been explained.  

The formalism is now ready for application to the SSM, and that will form the subject of the third paper in this series \ci{cybersusyIII}.

\appendix

\section{Conventions, Lorentz metric and Weyl spinors}
\la{fggegghetgrege}

\subsection{Lorentz metric and $\s$ matrices}

	The Lorentz metric is defined by the relation:
\begin{equation} x_{\mu} x^{\mu} = \eta_{\mu \nu}
x^{\mu} x^{\nu}  =  - x_{0}^{2} + x_{1}^{2} +
x_{2}^{2} + x_{3}^{2} \end{equation}
	The hermitian $2\times 2$  sigma matrices are
defined as usual:
\be \s_{1} = \left( \ba{cc} 0 &  1 \\ 1 &  0 \\ \ea
\right) \ee
\be \s_{2} = \left( \ba{cc} 0 &  - i\\  i &  0 \\
\ea
\right) \ee
\be \s_{3} = \left( \ba{cc} 1 &  0 \\ 0 & -1 \\ \ea
\right) \ee and the two dimensional unit matrix is
of course:
\be 1 = \left( \ba{cc} 1 &  0 \\ 0 & 1 \\ \ea
\right). \ee 
These satisfy the relations:
\be
\sum_{i=1}^3\s_{\a}^{i \b} \s_{\g}^{i \d}  = 
2 \d_{\a}^{\d}
 \d_{\g}^{\b}
-
 \d_{\a}^{\b}
 \d_{\g}^{\d}
\ee 
\be
\s_{\a}^{i \b} \s_{\b}^{j \g}  =
 \d^{ij}
 \d_{\a}^{\g}
+
i \e^{ijk} 
\s_{\a}^{k \g}
\; i = 1,2,3
\ee 
To develop Weyl spinors, we change notation a bit.

We take:
\be
\s^{0}_{\a \dot{\b}} =  -
\s_{0 \a \dot{\b}} = ( \bf{1})_{\a \dot \b} 
\ee and we let the other sigma matrices be given
by (i=1,2,3)
\be
\s^i = \s_i = (\s_{i})_{\a \dot \b}
\ee Let us summarize these definitions in the form:
\be
\s^{\m}_{ \a \dot \b} = ( 1,  \s^i )_{\a \dot  \b}
\ee The complex conjugate matrices are defined by:
\be (\s^{\m}_{\a \dot \b})^* =
\ov{\s}^{\m}_{\dot \a \b} =
\s^{\m}_{\b \dot \a}
\ee since the $\s$ matrices are hermitian.
Contrary to the usual convention, we do not
reverse the order of (anticommuting) spinors when
taking the complex conjugate. (Reversing the 
order of commuting spinors makes no difference of
course.)  Indices are raised and lowered as
follows:
\be
\y^{\a} = \varepsilon^{\a \b} \y_{\b}
\ee
\be
\y_{\a} =  - \varepsilon_{\a \b} \y^{\b}
\ee
\be
\ov{\y}^{\dot \a} = \varepsilon^{\dot \a \dot \b}
\ov{\y}_{\dot \b}
\ee
\be
\ov{\y}_{\dot \a} =  - \varepsilon_{\dot \a \dot
\b} \ov{\y}^{\dot \b}
\ee where the $\e$ tensors are real antisymmetric
matrices with:
\be
\varepsilon^{ \a  \g} 
\varepsilon_{ \b  \g}  =
\d^{\a}_{\b}
\ee
\be
\varepsilon^{\dot \a \dot \g} 
\varepsilon_{\dot \b \dot \g}  =
\d^{\dot \a}_{\dot \b}
\ee where  $\d^{\a}_{ \b} = 1 $ if $\a = \b $ and
$\d^{\a}_{ \b} = 0 $ if
$\a \neq \b$ (Same for 
$\d^{\dot \a}_{\dot \b}$). We can write this in
the form:
\be
\d^{\a}_{\b} = ({\bf 1})^{\a}_{\b}
\ee We take:
\be
\e^{\a \b} =  i (\s_2)^{\a \b}
\ee
\be
\e_{\a \b} =  i (\s_{ 2})_{ \a \b}
\ee Using this rule for raising and lowering
indices, we have:
\be
\ov{\s}^{\m \dot \a \b}  =
\e^{ \dot \a \dot \d}
\e^{ \b \g}
\ov{\s}^{ \m}_{\dot  \d \g}  = ( 1, - \s^i )^{\dot
\a \b}
\ee so that: 
\be
\ov{\s}_{\m}^{ \dot \a \b}  = - ( 1,  \s^i )^{\dot
\a \b}
\la{-sigma}
\ee where we use the relations:
\be
\s_2 (\s_i)^* \s_2  = -
\s_i
\ee

It is easy to check that the sigma matrices
satisfy the following relations:
\be
\s_{i} \s_{j} = \d_{ij} {\bf 1} +  i
\varepsilon_{ijk} \s_{k}
\ee which results in a number of other relations
such as:
\be
\s^{\m}_{\a \dot{\a}}
\ov{\s}^{\n \dot{\a} \b } +
\s^{\n}_{\a \dot{\a}}
\ov{\s}^{\m \dot{\a} \b } =
 - 2 \eta^{\m \n} \d^{\b}_{\a}
\ee
\be
\s^{\m}_{\a \dot{\b}}
\ov{\s}_{\m}^{ \dot{\g} \d } = - 2 \d^{\d}_{\a}
 \d^{\dot \g}_{\dot \b}
\ee We define:
\[
\s^{\m \n}_{\a \b} = 
\s^{\m \n}_{\b \a} =  - \s^{\n \m}_{\a \b} = 
\fr{1}{2} [
\s^{\m}_{\a \dot{\g}} 
\ov{\s}^{ \n \dot{\g}}_{\; \;\;  \b } -
\s^{\n}_{\a \dot{\g}} 
\ov{\s}^{ \m \dot{\g}}_{\; \;\;  \b }]
\]
\be
 = 
\fr{1}{2} [
\s^{\m}_{\a \dot{\g}}
\s^{\n}_{\b \dot{\d}}  -
\s^{\n}_{\a \dot{\g}}
\s^{\m}_{\b \dot{\d}}  ]
\varepsilon^{\dot{\g} \dot{\d} }
\ee and 
\[
\ov{\s}^{\m \n \dot{ \a} \dot {\b}} = 
\ov{\s}^{\m \n \dot{ \b} \dot {\a}} = 
-\ov{\s}^{\n \m \dot{ \a} \dot {\b}} 
\]
\be = -
\fr{1}{2} [
\ov{\s}^{ \m \dot{\a} \g } 
\s^{ \n \d \dot{\b} }  -
\ov{\s}^{ \n \dot{\a} \g } 
\s^{ \m \d \dot{\b} }  ]\varepsilon_{\g \d }
\ee Then 
\be (\s^{0i})_{\a}^{\;\b}  = - (\s^i)_{\a}^{\;\b} 
\ee
\be (\s^{ij})_{\a}^{\;\b}  = - i \varepsilon^{ijk}
(\s^k)_{\a}^{\;\b} 
\ee Let us use the following shorthand:
\be A \cdot \s^{\m} \cdot \ov{B}  = A^{\a}
\s^{\m}_{\a \dot{\b}} \ov{B}^{\dot \b} , \; 
\ov{A} \cdot \ov \s^{\m} \cdot B  =
\ov{A}^{\dot \a} \ov \s^{\m }_{ \dot \a \b} B^{ \b}
\ee The following identities help to familiarize
the notation.  For commuting spinors:
\be A \cdot B = - B \cdot A = A^{\a} B_{\a}
\ee
\be
\ov{A} \cdot \ov{B} = - \ov{B} \cdot \ov{A} = 
\ov{A}^{\dot \a} \ov{B}_{\dot \a}
\ee
For anticommuting spinors:
\be
\c \cdot \y =  \y \cdot \c = \y^{\a} \c_{\a}
\ee
\be
\ov{\c} \cdot \ov{\y} =  \ov{\y} \cdot \ov{\c} =
\ov{\y}^{\dot \a} \ov{\c}_{\dot \a}
\ee
We use latin letters for commuting
spinors and greek letters for anticommuting ones.
Here is another use of this dot product:
\be (\s^{\m} \cdot \ov{\s}^{\n} \cdot \s^{\l})_{\a
\dot \d} = (\s^{\m})_{\a \dot \b}
(\ov{\s}^{\n})^{\dot \b \g} (\s^{\l})_{\g \dot \d}
\ee 
It should be remembered that since
$\varepsilon_{\a \b}$ is antisymmetric, one gets:
\be
\c_{\a} \y^{\a} = - \c^{\a} \y_{\a} 
\ee Formulae involving products of these invariant
tensors can  be reduced using the basic relations:
\be
\s^{\m} \cdot \ov{\s}^{\n} \cdot \s^{\l} +
\s^{\l} \cdot \ov{\s}^{\n} \cdot \s^{\m} = 2
\eta^{\m \l} \s^{\n}  - 2 \eta^{\m \n} \s^{\l}  -
2 \eta^{\l \n} \s^{\m} 
\ee		
\be
\s^{\m} \cdot \ov{\s}^{\n} \cdot \s^{\l} -
\s^{\l} \cdot \ov{\s}^{\n} \cdot \s^{\m} = - 2i
\varepsilon^{\m \n \l \r} \s_{\r}
\ee where we define:
\be
\varepsilon^{0ijk} = 
\varepsilon^{ijk} 
\ee Similarly one gets:
\be (A \cdot \s^{\m} \cdot \ov{B} )^* = B \cdot
\s^{\m} \cdot \ov{A}  =
\ov{A} \cdot \ov{\s}^{\m} \cdot B 
\ee and in particular
\be (A \cdot \s^{\m} \cdot \ov{A} )^* = A \cdot
\s^{\m} \cdot \ov{A} =
\ov{A} \cdot \ov{\s}^{\m} \cdot A
\ee is a real quantity. The Fierz identity takes
the form:
\be A^{\a} \s^{\m}_{\a \dot \b}
 \ov{B}^{\dot \b} C^{\g} \s_{\m \g \dot \d}
\ov{D}^{\d}  = - 2 A^{\a} C_{\a} \ov{B}^{\dot \b}
\ov{D}_{\dot \b}
\ee or
\be A \cdot \s^{\m}\cdot  
\ov{B} C \cdot  \s_{\m}
\cdot \ov{D} = - 2 A\cdot  C \; \ov{B}
\cdot  \ov{D}
\ee for commuting spinors, with appropriate change
of sign for  the anticommuting case.
 
Note that since
the rule for raising is
\be 
A^{\a}=\ve^{\a \g} A_{\g}
\ee
and the rule for lowering is
\be 
A_{\b}=A^{\d}\ve_{\d \b}  = - \ve_{\b \d} A^{\d}
\ee
we have
\be 
\ve^{\a \g} \ve_{\b \g}
=
 \ve_{\b}^{\;\; \a}
=
- \ve_{\;\; \b}^{ \a}
=
 \d_{\;\; \b}^{ \a}
=
\d_{\b}^{\;\; \a}
=
\d_{\b}^{\a}
\ee

So this is consistent with the conventions.

\subsection{Spacetime coordinates}

Now define
\be
x^{\m}=
 x^{\a \dot \b}\s^{\m}_{\a \dot \b} 
\ee
\be
x_{\m}=
 x_{\a \dot \b}\s_{\m}^{\a \dot \b} 
\ee
Using the basic relations
\be
 \s^{\m}_{\a \dot \b} 
 \s^{\n \g \dot \b} 
+ \s^{\n}_{\a \dot \b} 
 \s^{\m \g \dot \b} 
= -2\h^{\m\n} \d^{\g}_{\a}
\ee
\be
\s^{\m}_{\a \dot \b} 
 \s^{\n \a \dot \b} 
= -2\h^{\m\n} 
\ee
and
\be
\s_{\m}^{\g \dot \d}  \s^{\m}_{\a \dot \b} 
= - 2 \d_{\dot \b}^{\dot \d} 
\d_{\a}^{\g} 
\ee
we see that the inverse is
\be
 x_{\a \dot \b}
=
- \fr{1}{2}
 x_{\m} \s^{\m}_{\a \dot \b} 
\ee
\be
 x^{\a \dot \b}
=
- \fr{1}{2}
 x^{\m} \s_{\m}^{\a \dot \b} 
\ee

Here is the derivation
\be
 x_{\a \dot \b}
=
a_1 x_{\m} \s^{\m}_{\a \dot \b} 
=
a_1 x_{\g \dot \d}\s_{\m}^{\g \dot \d}  \s^{\m}_{\a \dot \b} 
=
a_1 x_{\g \dot \d}
\lt (
-2 \d^{\g}_{\a} 
\d^{ \dot \d}_{ \dot \b} 
\rt )
\eb
= -2
a_1 
 x_{\a \dot \b}
\eb
\Ra a_1 = - \fr{1}{2}
\ee

\subsection{Derivatives}
So
\be
\pa_{\a \dot \b} 
=
\fr{\pa}{\pa x^{\a \dot \b} }
=
\fr{\pa}{\pa x^{\m} }
\fr{\pa x^{\m} }{\pa x^{\a \dot \b}}  
\eb
=
\fr{\pa}{\pa x^{\m} }
\fr{\pa (x^{\g \dot \d} \s^{\m}_{\g \dot \d})}{\pa x^{\a \dot \b}}  
=
\fr{\pa}{\pa x^{\m} }
  \s^{\m}_{\a \dot \b}    
=
\pa_{\m} 
  \s^{\m}_{\a \dot \b}    
\ee
which summarizes as
\be
\pa_{\a \dot \b} 
=\pa_{\m} 
  \s^{\m}_{\a \dot \b}    
\ee
The inverse is
\be
\pa_{\m} 
=
\fr{\pa}{\pa x^{\m} }
=
\fr{\pa x^{\a \dot \b} }{\pa x^{\m}}  
\fr{\pa}{\pa x^{\a \dot \b} }
\eb
=
\fr{\pa \lt ( \fr{-1}{2} x^{\n} \s_{\n}^{\a \dot \b} \rt )
}{\pa x^{\m}}  
\fr{\pa}{\pa x^{\a \dot \b} }
=
 \fr{-1}{2} \s_{\m}^{\a \dot \b} 
\fr{\pa}{\pa x^{\a \dot \b} }
\ee
And also 
\be
x_{\a \dot \b} 
y^{\a \dot \b} 
=
x_{\m} \s^{\m}_{\a \dot \b} 
y_{\n} \s^{\n \a \dot \b} 
= - 2 
x_{\m} 
y^{\m}
\ee

\subsection{Product of Derivatives}

\be
\pa_{\a \dot \b} 
\pa^{\a \dot \g} 
=\pa_{\m} \pa_{\n} 
  \s^{\m}_{\a \dot \b}    
  \s^{\n \a \dot \g}    
\eb
=\pa_{\m} \pa_{\n} 
\fr{1}{2}  \lt (
\s^{\m}_{\a \dot \b}    
  \s^{\n \a \dot \g}    
+\s^{\n}_{\a \dot \b}    
  \s^{\m \a \dot \g}    
\rt )
\eb
=\pa_{\m} \pa_{\n} 
\fr{1}{2}  \lt (
- 2 \h^{\m \n} 
\d_{\dot \b}^{\dot \g}    
\rt )
= \D    
\d_{\dot \b}^{\dot \g}    
\ee
Similarly:
\be
\pa_{\a \dot \b} \pa^{\g \dot \b} 
=
\D \d_{\a }^{\;\; \g} 
\ee
\be
\ve^{\g \a}  \pa_{\a \dot \b} 
=
\pa^{\g}_{\;\; \dot \b} 
\ee
\be
\pa^{\a \dot \b} \ve_{\a \g}  
=
\pa_{\g}^{\;\; \dot \b} 
\ee
\be
\ve_{\a \g}  
\ve^{\d \g}  
=
\d_{\a}^{\d}  
\ee

\subsection{Equations of motion and mass and the d'Alembertian}
\la{tqetqergoprthhop}

That means that in fact
\be
\pa_{\a \dot \b} \pa^{\a \dot \b} 
=
2 \D  
\ee
and so
\be
\D   = \fr{1}{2}
\pa_{\a \dot \b} \pa^{\a \dot \b} 
\ee
In the present paper, we have used the notation:
\be X = \fr{\D}{m^2}
\ee
So the d'Alembertian operator has the form:
\be
\pa_{\m} 
\pa^{\m} 
\eb
=
\lt (
\fr{-1}{2}
\pa_{\a \dot \b} 
\s_{\m}^{\a \dot \b} 
\rt )
\lt (
\fr{-1}{2}
\pa^{\g \dot \d} 
\s^{\m}_{\g \dot \d} 
\rt )
=
\fr{-1}{2}
\pa_{\a \dot \b} 
\pa^{\a \dot \b} 
\ee
and so we have
\be
\fr{1}{2}
\pa_{\a \dot \z} 
\pa^{\a \dot \z} 
=
\D
=-
\pa_{\m} 
\pa^{\m} 
\ee
Now the equation of motion for a particle  involves the expression:
\be
 \pa_{\m}\pa^{\m}
\ra - p_{\m} p^{\m}
=  m^2
\ee
if it is on shell.
So the correct signs are
\be
p_{\m} p^{\m}
+  m^2
=- p_{0}^2 +  p_{i}
p_{i}
+  m^2
\ee
for a propagator
and
\be
\D +   m^2
= -\pa_{\m}\pa^{\m}
+  m^2
= \pa_{0}\pa_{0}- \pa_{i}\pa_{i}
+  m^2
\ra - p_{0}^2 + p_{i}
p_{i}
+  m^2
\ee
for an equation of motion.

$X$ is a  parameter defined as follows:

\be
X = \fr{\D}{m^2} =  \fr{- \pa^{\m}\pa_{\m} }{m^2} 
= \fr{ \pa_{0}\pa_{0}-\pa_{i}\pa_{i} }{m^2}
 \ra    \fr{- p_0^2 + p_i p_i }{m^2}   
\ee
$X$ appears when we solve for the propagators which are the inverse of the kinetic terms. It is defined to be dimensionless.
Masses arise in the theory for values of X for which the denominators of propagators go to zero.  Thus if we have a propagator 
\be
\fr{1}{m^2(X+ X_0)}=\fr{1}{ - p_0^2 + p_i p_i  +X_0 m^2} 
\ee
then there is a mass at a positive value of $X_0$, and the value  of $X$ is for that mass is negative.

\listoftables

\tableofcontents

\end{document}